\documentclass[twocolumn]{aastex62}
\usepackage[utf8]{inputenc}
\usepackage{hyperref}
\usepackage{silence}
\usepackage{graphicx}
\usepackage[caption=false]{subfig}

\usepackage{float}
\usepackage{color,soul}

\usepackage{wrapfig}
\usepackage{bm}
\usepackage{amsmath}

\usepackage{fancyhdr}
\usepackage{wasysym}
\usepackage{natbib}
\usepackage{color, colortbl}
\usepackage{wrapfig}

\begin{document}
\defcitealias{Hill2013}{H13}
\defcitealias{Taylor2009}{T09}

\newcommand{\RMhvc}{\ensuremath{\textrm{RM}_\mathrm{HVC}}}

\title{Constraining the Magnetic Field of the Smith High Velocity Cloud using Faraday rotation}

\author{S. K. Betti}
\affiliation{Department of Physics and Astronomy, Haverford College, Haverford, PA 19041, USA }
\affiliation{Department of Astronomy, University of Massachusetts, Amherst, MA 01003, USA}

\author{Alex S. Hill}
\affiliation{Department of Physics and Astronomy, Haverford College, Haverford, PA 19041, USA }
\affiliation{Department of Physics and Astronomy, University of British Columbia, Vancouver, BC V6T 1Z1, Canada} 
\affiliation{Space Science Institute, Boulder, CO 80301, USA}
\affiliation{Dominion Radio Astrophysical Observatory, National Research Council, Kaleden, BC V0H 1K0, Canada} 

\author{S. A. Mao} 
\affiliation{Max Planck Institute for Radio Astronomy, Auf dem H\"{u}gel 69, D-53121 Bonn, Germany}

\author{B. M. Gaensler}
\affiliation{Dunlap Institute for Astronomy \& Astrophysics, University of Toronto, Toronto, ON, M5S
3H4, Canada}

\author{Felix J. Lockman}
\affiliation{Green Bank Observatory, P.O. Box 2, Green Bank, WV 24944, USA}

\author{N. M. McClure-Griffiths}
\affiliation{Research School of Astronomy and Astrophysics, Australian National University, Canberra, ACT 2611, Australia}

\author{Robert A. Benjamin}
\affiliation{Department of Physics, University of Wisconsin-Whitewater, Whitewater, WI 53190, USA}

\correspondingauthor{S. K. Betti}
\email{sbetti@umass.edu}

\singlespace 
\begin{abstract}
The Smith Cloud is a high velocity cloud (HVC) with an orbit suggesting it has made at least one passage through the Milky Way disk.  
A magnetic field found around this cloud has been thought to provide extra stability as it passes through the Galactic halo. 
We use the Karl G. Jansky Very Large Array to measure Faraday rotation measures (RMs) towards 1105 extragalactic background point sources behind and next to the Smith Cloud to constrain the detailed geometry and strength of its magnetic field.   
The RM pattern across the cloud gives a detailed morphology of the magnetic field structure which indicates a field draped over the ionized gas and compressed at the head of the cloud.  We constrain the peak line-of-sight magnetic field strength to $\gtrsim$+5 $\mu$G and create a model of the magnetic field to demonstrate that a draped configuration can qualitatively explain the morphology of the observed RMs. 
\end{abstract}
\doublespace

%-------------------------------------------------------------------------------------------------------------------------

\section{Introduction} \label{Sec1}

High velocity clouds (HVCs) are gas clouds within the halo of the Milky Way with velocities inconsistent with the rotation of the galactic disk.  
HVC candidates have been observed in M31 and M33, but most have been seen around the disk of the Milky Way \citep{Wright1979, Wakker1997}. Many, though not all, of these clouds are falling towards the disk, potentially providing material for future star formation and driving galaxy evolution \citep{Putmanreview}.  
  
\begin{figure*}[tb] 
\centering
\includegraphics[width = 0.9\linewidth]{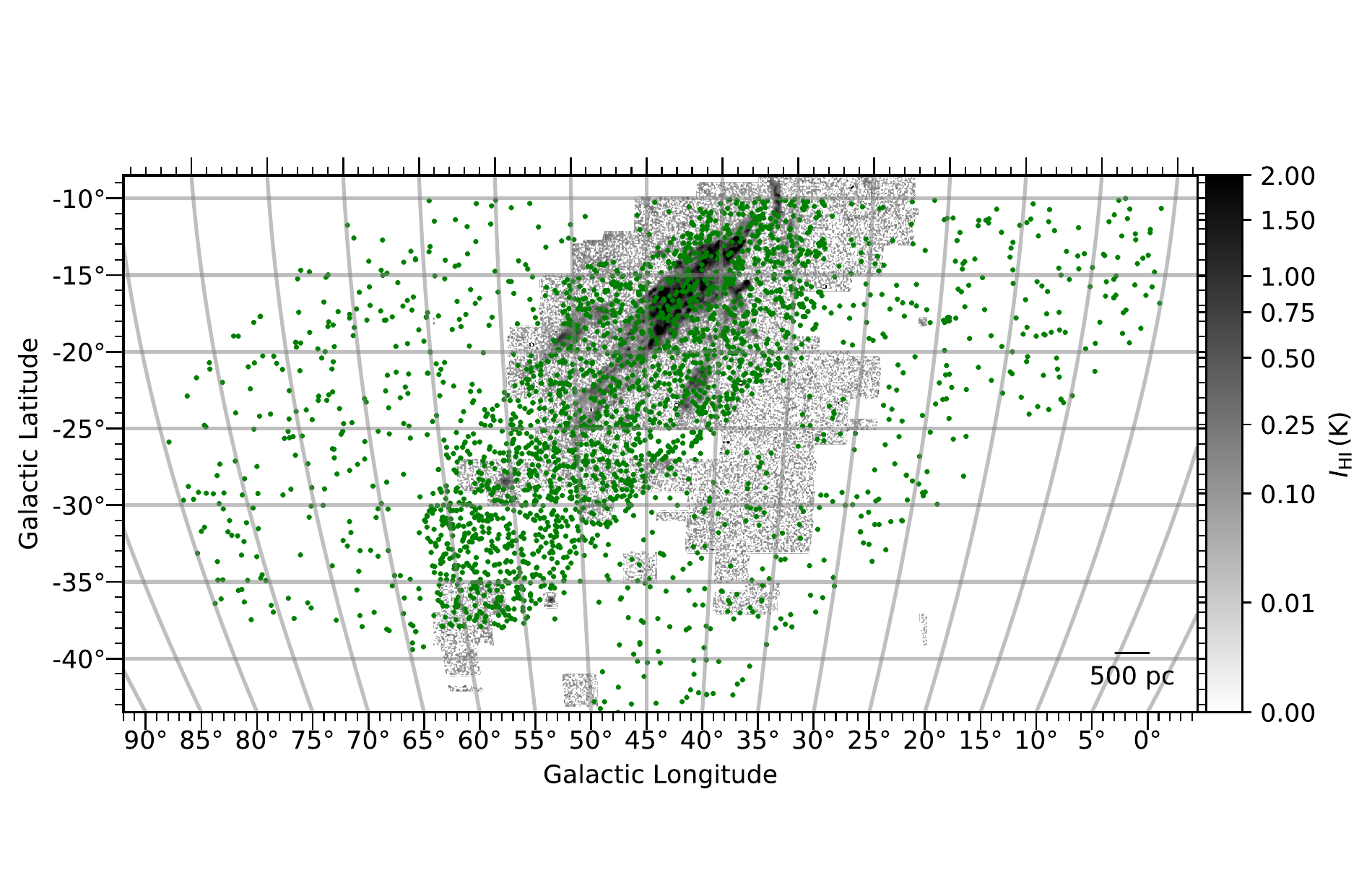}
\caption{2847 extragalactic sources selected from the NVSS catalog (chosen as described in Section~\ref{Sec2}) overlaid on GBT \ion{H}{1} emission in the $v_\mathrm{GSR} =+247$ km s$^{-1}$ channel.} 
\label{RM-sources}
\end{figure*}

The Smith Cloud \citep{Smith1963, Bland-Hawthorn1998} is an HVC in which we clearly see the interaction at the disk-halo interface due to its large projected size (see Figure \ref{RM-sources}) and proximity to the Milky Way \citep{Lockman2008}. 
The Smith Cloud has been measured to be 12.4 $\pm$ 1.3 kpc from the Sun \citep{Wakker2008, Putman2003, Lockman2008} and $3$ kpc below the Galactic plane.  It has a cometary morphology with a bright knot at $(l, b) = (395\arcdeg, -13\arcdeg)$ and more diffuse H~{\sc i} emission trailing away from the plane, which \citet{Lockman2008} interpreted as evidence that the cloud is moving towards the disk with a speed $v_z = +73 \textrm{ km s}^{-1}$ and a 3D velocity of $296 \textrm{ km s}^{-1}$, calling the bright knot the ``head'' and the more-diffuse emission the ``tail''. Based on this assumption as well as the gradient in radial velocity across the cloud, \citet{Lockman2008} and \citet{Nichols2009} each calculated orbits for the cloud in which it passed through the disk $\approx 70$~Myr ago. There are Rayleigh-Taylor instabilities in the \ion{H}{1} observations which are morphologically similar to those seen in hydrodynamic simulations of clouds falling into the disk \citep{Armillotta2017}, further evidence that the cloud is falling towards the disk. However, \citet{Tepper-Garcia2018} suggest that the cloud may be a dark matter ``streamer'' currently moving away from the disk. There is no proper motion measurement of the cloud.  Though there is no detectable stellar population \citep{Stark2015}, a high sulfur abundance of [S/H] = 0.53$^{+0.21}_{-0.15}$ has been found suggesting an origin from the outer disk of the Galaxy \citep{Fox2016}.

According to hydrodynamical simulations \citep{Heitsch2009}, a passage of more than 10 kpc through the Galactic halo should strip an HVC of its neutral hydrogen content.  Though the Smith Cloud has traveled more than $10$ kpc and the \ion{H}{1} of the Smith Cloud is disrupted, an \ion{H}{1} mass of $> 10^{6} M_{\astrosun}$ has been measured in coherent structures \citep{Lockman2008} with an additional $\sim 10^{6} M_\odot$ in ionized gas \citep{Hill2009}.

Two-dimensional magnetohydrodynamic (MHD) simulations) of HVC environs suggest that a magnetic barrier of a few $\mu$G between the HVC and the hot surrounding medium could reduce any disruption and gas stripping \citep{Konz2002}. If this magnetic barrier is an ambient field that has been swept up as the cloud travels toward the Galactic disk, then 3D MHD simulations \citep{Gronnow2017} suggest the cloud is within $10$ kpc of the disk.  \citet{Gronnow2018} found that in MHD simulations for strong magnetic fields with sub-Alfvenic flows, the field can suppress condensation and forms Kevin-Helmholtz instabilities and flattening in the wake of the cloud. For super-Alfvenic flows, spherical objects interacting with magnetized plasma flow cause a perpendicular background magnetic field to ``drape'' around the object, resulting in a flattening and funneling of the cloud \citep{Mac-LowMcKee:1994,JonesRyu:1996,Dursi2008,Romanelli2014}. Models of Smith Cloud-like HVCs show evidence of similar draping \citep{GalyardtShelton:2016}.

One way to estimate the magnetic field in an ionized cloud is through measurements of the Faraday rotation of polarized radio continuum sources that lie behind it.  \citet{McClureGriffiths2010}, \citet[][hereafter H13]{Hill2013}, and  \citet{KaczmarekPurcell:2017} used this technique, taking rotation measures (RMs) from the catalog of \citet[][hereafter T09]{Taylor2009} or from new observations and estimates of the ionized gas density and distribution from H-alpha measurements using the Wisconsin H-Alpha Mapper (WHAM; \citealt{Haffner2003, Haffner2010}) to measure the magnetic field of an HVC in the leading arm of the Magellanic System, the Smith Cloud, and the Magellanic Bridge respectively.  \citetalias{Taylor2009} derived RMs from the NRAO VLA Sky Survey \citep[NVSS;][]{Condon1998}, which imaged the entire northern sky with declination $\delta > -40{\arcdeg}$ at 1.4 GHz in continuum total intensity and linear polarization.  The \citetalias{Taylor2009} catalog contains RMs for $\sim$1 point source deg$^{-2}$ with polarized intensities ($PI$) $> 0.4$ mJy.  \citetalias{Hill2013} derived a magnetic field $\geq 8 \mu \mathrm{G}$ in gas associated with the Smith Cloud.
The relatively sparse spatial sampling of the \citetalias{Taylor2009} catalog did not allow strong constraints on the structure of the magnetic field of the Smith Cloud.  For this reason we have undertaken more detailed radio continuum observations of the area on and around the Smith Cloud and derived RMs for fainter continuum sources than considered by \citetalias{Taylor2009}.  

In this paper, we present RMs for 1105 sources in order to gain a more detailed knowledge of the morphology of the magnetic field of the Smith Cloud.  
In Section \ref{Sec2}, we discuss our observations of Faraday rotation, while in Section \ref{Sec3} we describe our results.  We estimate magnetic field strengths in different regions of the cloud in Section 4, while in Section 5, we used a toy model of a draped magnetic field to reproduce the observe rotation measure pattern across the Smith Cloud.  Finally, we summarize the paper in Section \ref{Sec6}.

%-------------------------------------------------------------------------------------------------------------------------

\section{Data} \label{Sec2}

To greatly increase the number of sight lines through the Smith Cloud with magnetic field measurements over what was possible with the \citetalias{Taylor2009} NVSS RM catalog and to obtain more reliable individual RM measurements, we observed 2847 extragalactic sources behind the Smith Cloud using the upgraded Karl G. Jansky Very Large Array (JVLA).  The sources chosen are bright polarized radio point sources; most are active galactic nuclei.  

We chose polarized point sources from the NVSS catalog with $PI > 0.1$~mJy; there are approximately six times as many sources per square degree which meet this criterion as there are sources with $PI > 0.4$~mJy, the cutoff used in the \citetalias{Taylor2009} NVSS RM catalog.   Within $10{\arcdeg}$ of the major axis of the cloud, we observed all of these sources, totaling around 2100 sources (``on cloud").  This high density was chosen to measure the structure of the magnetic field in the cloud on $\sim$degree scales. In a larger area (between $10{\arcdeg}$ and $20{\arcdeg}$ from the major axis), we randomly chose $1/6$ of the $\approx5400$ available sources with $PI > 0.1$~mJy, $\sim$750 sources 
(``off cloud") (see Figure \ref{RM-sources}).  We use these ``off cloud" sources to measure the foreground Milky Way contribution; this contribution must be subtracted out to determine the RMs associated with the Smith Cloud.

%-------------------------------------------------------------------------------------------------------------------------

\subsection{Observations} 
We observed 2847 sources using the JVLA between 2012 October 19 $-$ December 8.  The observations covered $1-2$ GHz in the A array configuration.  With the 1.3$\arcsec$ resolution of the A configuration (compared to 46$\arcsec$ resolution in the D configuration used for the NVSS), approximately 2000 sources are resolved in our observations.   
We chose the exposure time to achieve a polarization signal-to-noise ratio of $\sim$10 across the 1 GHz-wide band based on the NVSS $PI$ accounting for an expected loss of $\approx 40\%$ of the band due to radio frequency interference (RFI). Exposure times ranged from $10-30$~s depending on the polarized intensity from NVSS. With short exposure times and numerous sources, slew time was a significant fraction of the total observing time.  The {\tt MIRIAD} \citep{MIRIAD} task {\tt atmos} was used to solve the ``traveling salesman" problem, optimizing the source in order to minimize the slew time to 1 s per slew, on average, placing each NVSS source on axis. 

The standard flux calibrator 3C286 was observed before and 3C48 was observed after each observing run to allow for amplitude, bandpass, flux, and polarization angle calibration.  Leakage correction was calibrated using NVSS J235509+495008, which we assumed to be unpolarized.
We observed 1024 channels in 16 spectral windows with 1~MHz channel widths. The data was reduced using the Common Astronomy Software Applications \citep[CASA;][]{McMullin2007}.  With the {\tt rflag} option in CASA, radio frequency interference (RFI) was flagged; the $1.20-1.25 \textrm{ GHz}$ and $1.55-1.65 \textrm{ GHz}$ frequency ranges were mostly unusable. We then rebinned to $8 \textrm{ MHz}$ channels and applied the amplitude, bandpass, flux, polarization angle, and leakage calibration.

For each source, we made image cubes of Stokes parameters $I$, $Q$, $U$ using a cleaning threshold of $0.4~\rm{mJy}$, natural weighting, and a gain of 0.1.  The location of each source was found by using an IDL clump finding algorithm, FIND,\footnote{From the IDL Astronomy Users Library, \url{http://idlastro.gsfc.nasa.gov/ftp/pro/idlphot/find.pro}} in Stokes $I$ in order to find the corresponding Stokes $Q$ and $U$ values of the source, which were then used to compute the RM.  The clump which contained both the largest number of pixels and brightest flux was taken to be the source.  If no clump or RM was detected, the location of the source was then found by eye in Stokes $I$.     

%-------------------------------------------------------------------------------------------------------------------------

\subsection{Faraday Rotation}
Faraday rotation is a physical effect that depends on the polarized nature of radio emission and which allows us to measure the line-of-sight magnetic field in ionized gas. 
The polarized plane at wavelength $\lambda$ rotates by an angle 
\begin{equation} 
\Delta\chi = \text{RM} \, \lambda^{2}, 
\end{equation} 
where the RM is a measure of the integrated line-of-sight magnetic field, $B_{||}$, in ionized gas weighted by electron density, $n_{e}(s)$, over the line of sight,
\begin{equation} \label{eqn2}
\text{RM} = 0.812\int_{\text{source}}^{\text{observer}} \frac{n_{e}(s)}{\text{cm}^{-3}} \frac{B_{||}}{\mu\text{G}} \frac{ds}{\text{pc}} \ \,  \text{rad} \ \, \text{m}^{-2}.
\end{equation}
The sign and magnitude of the RM gives the direction and magnitude of the line-of-sight component of the magnetic field.  A positive RM is consistent with a line-of-sight magnetic field pointing towards the observer.  
We derive the electron density from the emission measure (EM) of H$\alpha$ from the interstellar medium (ISM) \citep{Hill2013, Mao2010} and describe the internal electron density within each clump as   
\begin{equation}
n_e = \sqrt{\frac{\text{EM}}{L_{H+}}}.
\end {equation}  
Here $L_{H+}$ is the path length occupied by ionized gas through the HVC found by $L_{H+} = fL$, where $f$ is the fraction of the total path length ($L$) through the HVC in which gas is present.  

\begin{figure*}[bt]    
       \begin{minipage}{.5\textwidth}
        \centering
        \includegraphics[width=\linewidth]{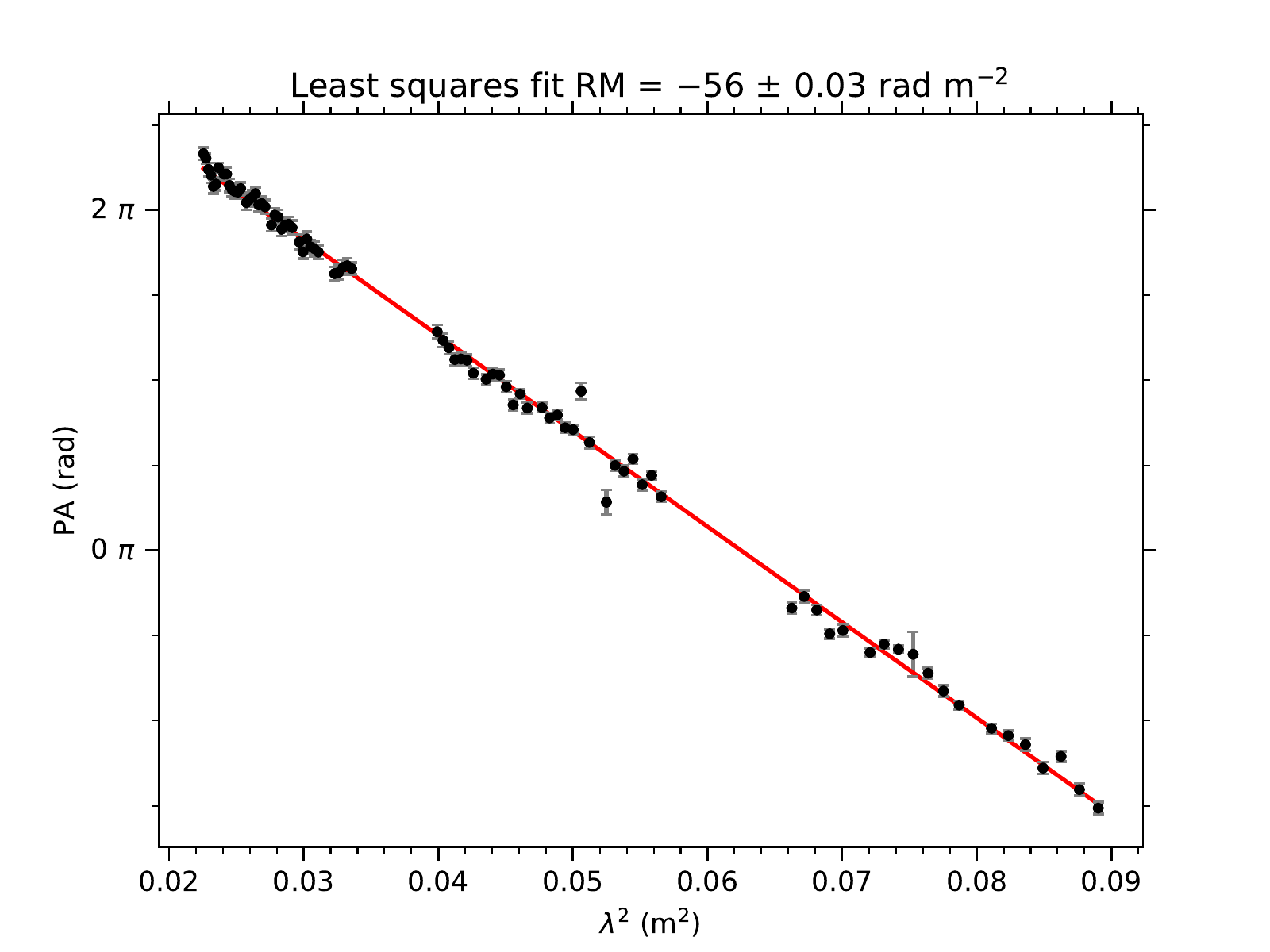}
    \end{minipage}%
    \begin{minipage}{0.5\textwidth}
        \centering
        \includegraphics[width=\linewidth]{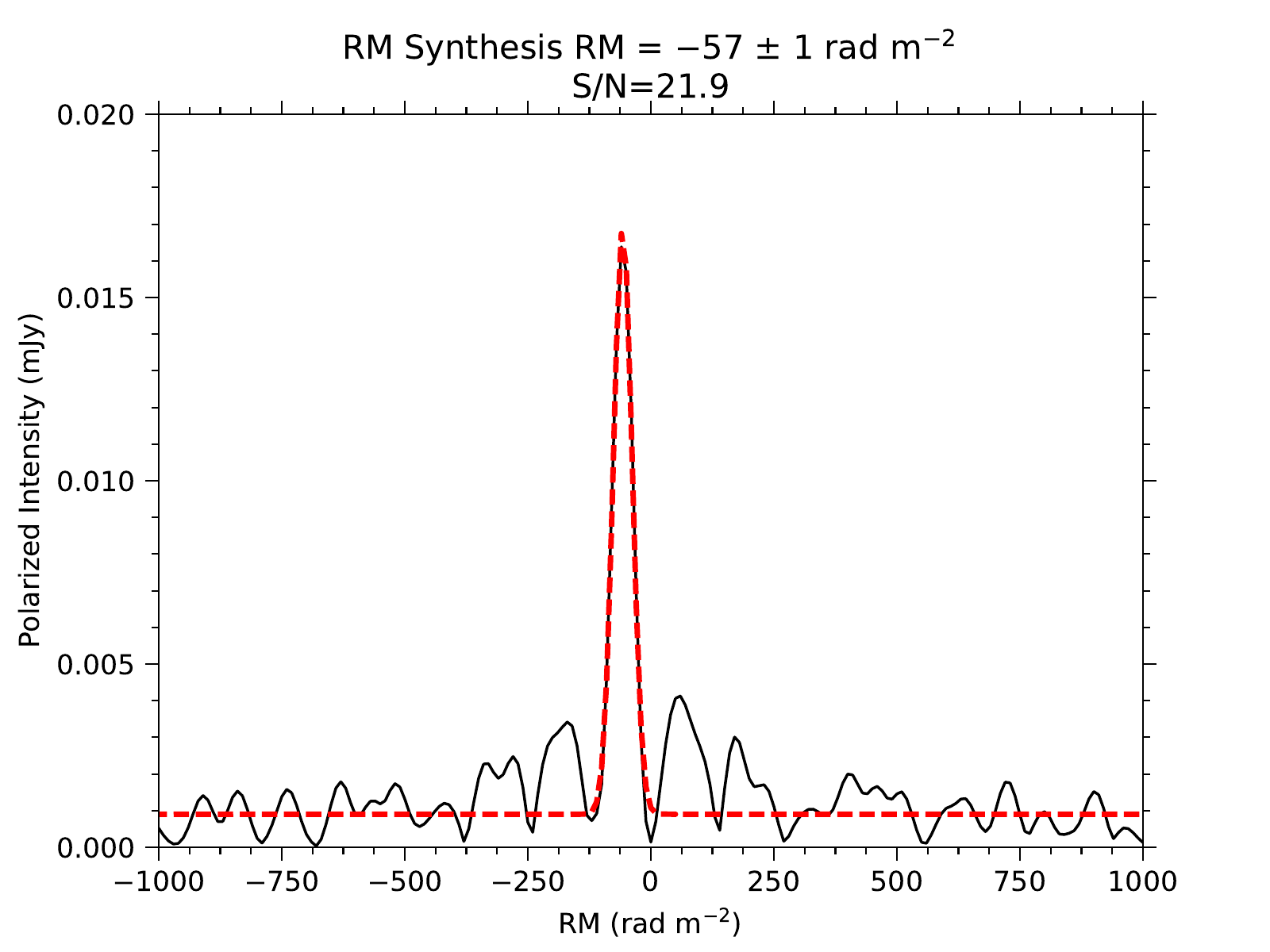}
    \end{minipage}
    \begin{minipage}{.5\textwidth}
        \centering
        \includegraphics[width=\linewidth]{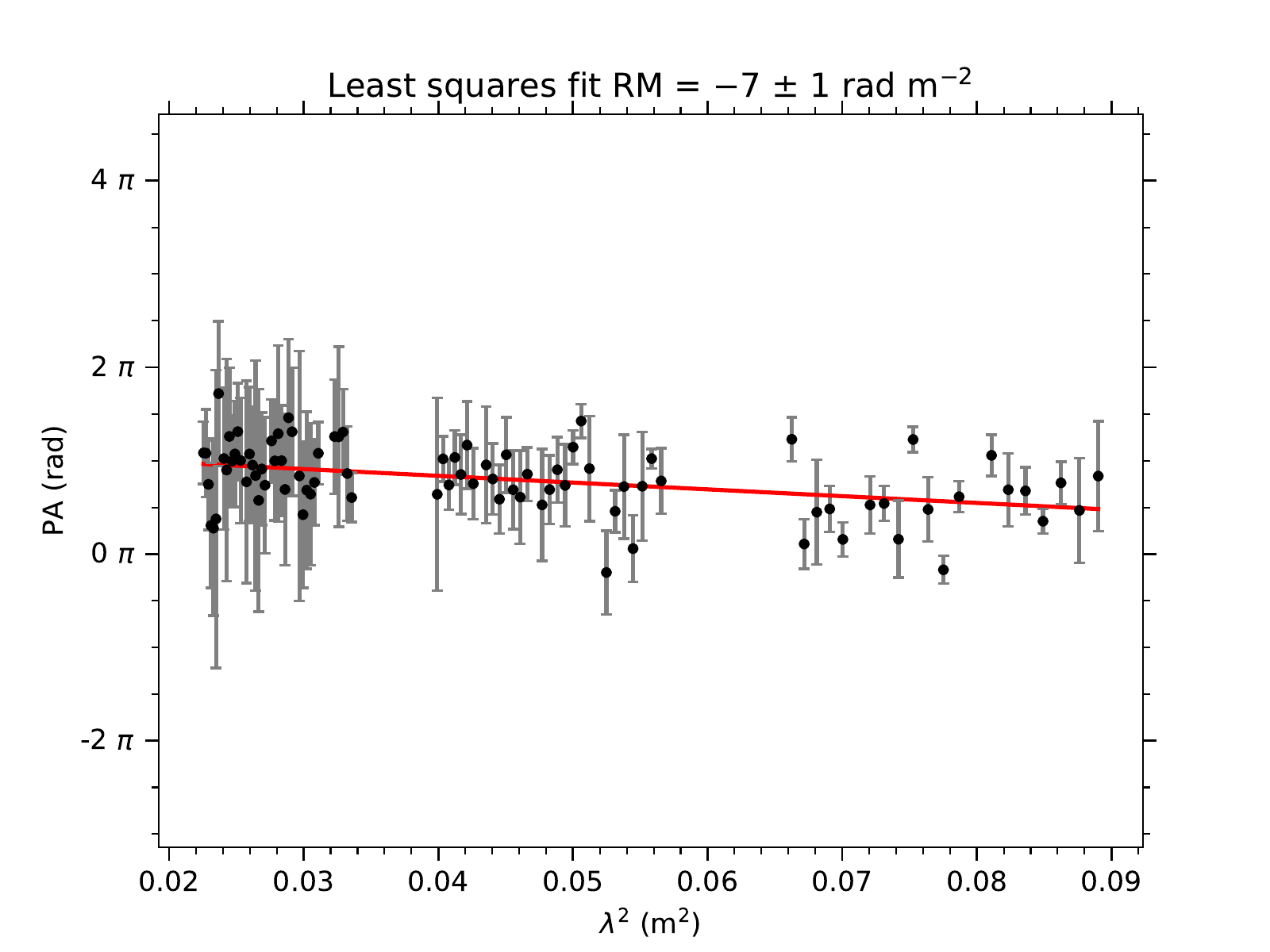}
    \end{minipage}%
    \begin{minipage}{0.5\textwidth}
        \centering
        \includegraphics[width=\linewidth]{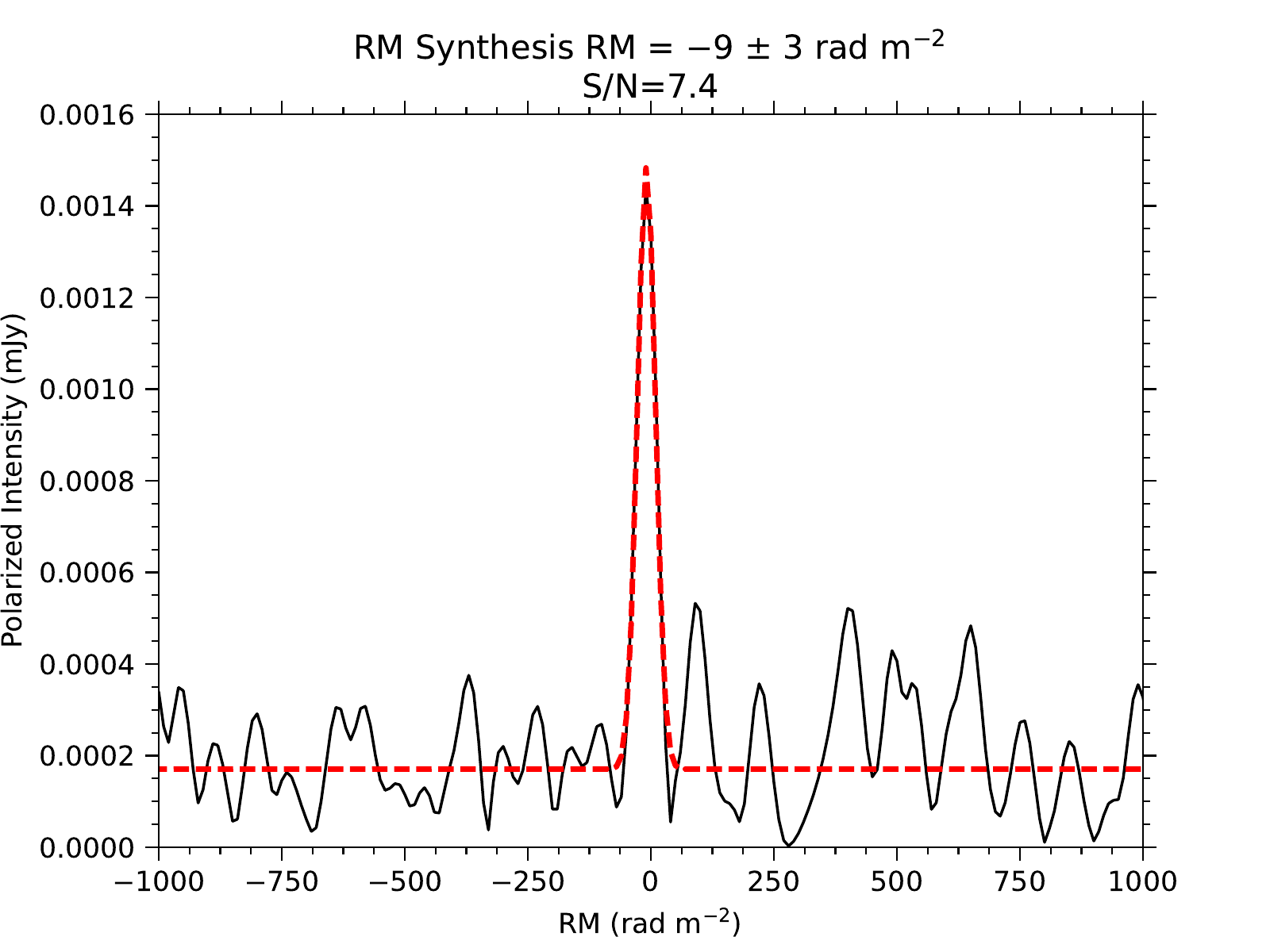}
    \end{minipage}
 \caption{Least-squares fit of the polarization angle and RM synthesis of sources ($l$, $b$) = (38.3428${\arcdeg}$, $-$17.6920${\arcdeg}$) (top) and ($l$, $b$) = (46.2540${\arcdeg}$, $-$29.4137${\arcdeg}$) (bottom) .  The left panel shows the traditional method, polarization angle as a function of wavelength squared, with a best fit line shown in red.  The RM is the slope of the line.  The right panel shows the Faraday dispersion spectrum, the amplitude of the linearly polarized intensity as a function of RM.  The fitted Gaussian is shown in red with the center of the Gaussian indicating the RM.} 
    \label{fig:polangleRM}
\end{figure*}

\begin{figure}[tb]
\centering
\includegraphics[width = 1.\linewidth]{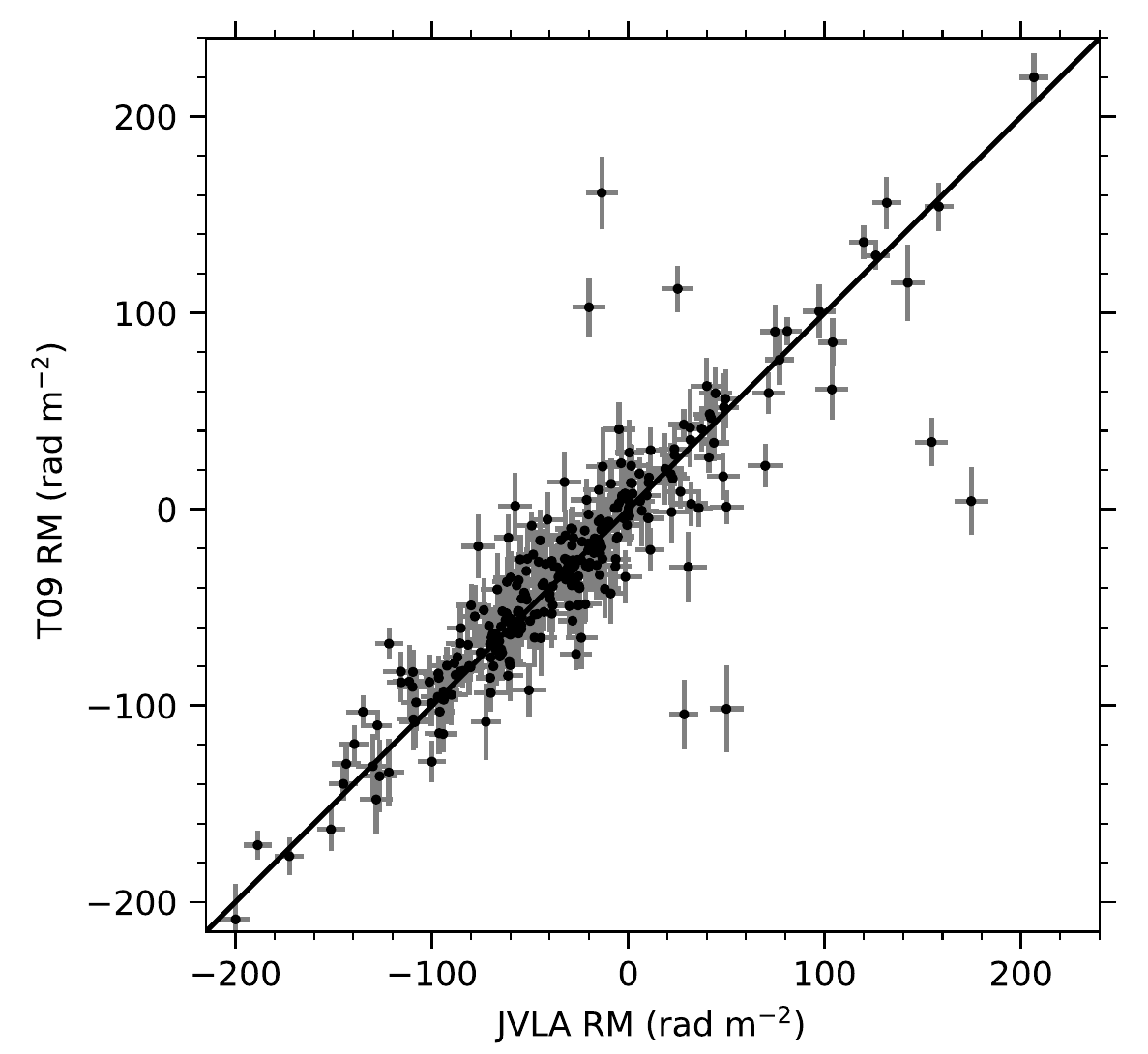}
\caption{Comparison of \citetalias{Taylor2009} RMs to their corresponding JVLA RMs.  In total, 258 significantly detected JVLA RMs had matches in the \citetalias{Taylor2009} NVSS RM catalog.  The line corresponds to RM$_{\text{T09}} =$ RM$_{\text{JVLA}}$.}
\label{figure:TvsVLA}
\end{figure}

%-------------------------------------------------------------------------------------------------------------------------

\subsection{RM computation}

\begin{deluxetable*}{cccccc@{ $\pm$ }lc@{ $\pm$ }lc}[tb]
\centering
\tablecaption{Rotation Measures 
\label{first10RM}
}
\tablehead{\colhead{$l$} &  \colhead{$b$} & \colhead{RA (J2000)} & \colhead{Dec (J2000)} & \colhead{S/N\tablenotemark{a}} & \multicolumn2c{RM} & \multicolumn2c{\RMhvc} & \colhead{Match\tablenotemark{b}} \\ \colhead{(deg)} & \colhead{(deg)} & \colhead{(deg)} & \colhead{(deg)} & \colhead{} & \multicolumn2c{(rad m$^{-2}$)} &  \multicolumn2c{(rad m$^{-2}$)} & \colhead{} }
\startdata
10.7671	&	$-$14.8100	&	286.7685	&	$-$26.1785	&	5.0	&	$-$61	&	5	&	$-$39.40	&	8.7	&	0	\\
10.7674	&	$-$15.7770	&	287.7615	&	$-$26.5574	&	9.7	&	$-$29	&	3	&	$-$10.60	&	7.5	&	0	\\
10.8719	&	$-$10.6613	&	282.6298	&	$-$24.3839	&	5.6	&	{$+$}70	&	5	&	$+$102.96	&	8.3	&	0	\\
11.0423	&	$-$18.4760	&	290.6755	&	$-$27.3360	&	5.0	&	\phn$+$6	&	5	&	$+$16.84	&	8.7	&	0	\\
11.6831	&	$-$13.8990	&	286.2352	&	$-$25.0012	&	6.1	&	$-$30	&	4	&	$-$5.48	&	8.1	&	0	\\
11.9586	&	$-$15.6490	&	288.1277	&	$-$25.4515	&	5.3	&	$+$30	&	1	&	$+$49.72	&	7.1	&	0	\\
12.9551	&	$-$10.7016	&	283.6159	&	$-$22.5470	&	4.0	&	$-$63	&	6	&	$-$27.85	&	9.5	&	0	\\
13.0453	&	$-$11.3240	&	284.2677	&	$-$22.7295	&	6.5	&	$-$102	&	4	&	$-$68.79	&	8.0	&	0	\\
13.1394	&	$-$16.3130	&	289.2911	&	$-$24.6649	&	5.9	&	$-$18	&	4	&	\phn$+$0.43	&	8.2	&	0	\\
13.3193	&	$-$15.0619	&	288.1029	&	$-$24.0142	&	6.2	&	$+$120	&	4	&	$+$142.47	&	8.1	&	0	\\
\enddata
\tablenotetext{a}{S/N refers to the signal to noise of the Faraday depth spectrum.}
\tablenotetext{b}{Match indicates if there is a counterpart in the \citetalias{Taylor2009} NVSS RM catalogue (1$-$yes, 0$-$no)}
\tablecomments{Table 1 is published in its entirety in the
machine readable format. A portion is shown here for guidance regarding its form and content.}
\end{deluxetable*}      

In order to obtain RMs for the sources, we follow the method of RM synthesis described by \cite{Mao2010} to determine the Faraday depth ($\phi$) of each source.  As we are observing Faraday rotation towards background point sources, the emitting region is entirely behind the Faraday-rotating medium we are interested in. We therefore treat the sources as Faraday simple and assume that $\phi = \, \mathrm{RM}$; \citet{MaMao:2018} investigate the Faraday complexity in our data.
With an observed bandwidth of $0.0225 \textrm{ m}^2 < \lambda^2 < 0.09 \textrm{ m}^2$, the expected FWHM of the rotation measure spread function \citep[RMSF;][]{Brentjens2005} is $\delta \phi \approx 56 \textrm{ rad m}^{-2}$.  

Due to RFI flagging and removal, there are gaps in the data. This causes sidelobes in the Faraday depth spectrum.  In order to obtain accurate RMs, we deconvolve the RMSF for the Faraday depth function using a version of the {\tt RMCLEAN} algorithm based on both the {\tt Clean\_RMS} \citep{Mao2010} and {\tt RMCLEAN} \citep{Heald2009} algorithms implemented in IDL by T. Robishaw (private communication).  
We use a gain factor of 0.1 and stopped {\tt RMCLEAN} at 5 times the polarized root mean squared (rms) level.  We show examples of the Faraday depth spectrum in the right panels of Figure \ref{fig:polangleRM}.  For comparison, the measured Faraday rotation from fitting the slope of $\chi$ versus $\lambda^2$ is given in the left panels of Figure \ref{fig:polangleRM}.

The value of RMs is found by fitting a Gaussian to the peak of the Faraday dispersion function.
We estimate the measurement uncertainty of the RMs as
\begin{equation}
\sigma_\mathrm{RM} = \frac{\sigma_{F(\phi)} \delta\phi}{ 2PI},
\end{equation}
where $\delta\phi$ is the FWHM of the observed RMSF (see equation 61 in \citealt{Brentjens2005}) and $\sigma_{F(\phi)}$ is the rms of the Faraday depth spectrum \citep{Schnitzeler2010, Schnitzeler2017, Mao2010}.

%-------------------------------------------------------------------------------------------------------------------------
\section{Results} \label{Sec3}

\begin{figure*}[bt]    
    \begin{minipage}{.5\textwidth}
        \centering
        \includegraphics[width=\linewidth]{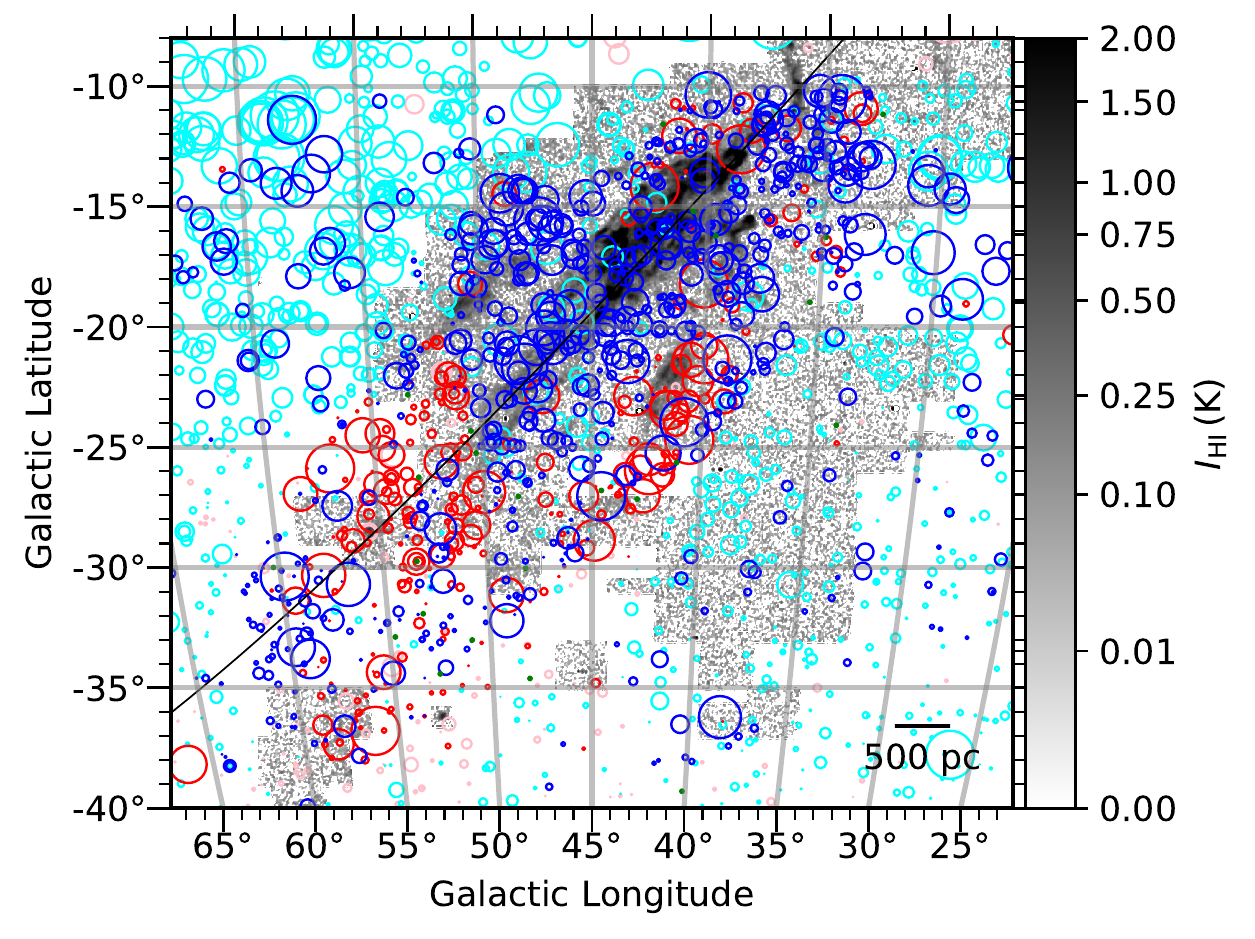}
    \end{minipage}%
    \begin{minipage}{0.5\textwidth}
        \centering
        \includegraphics[width=\linewidth]{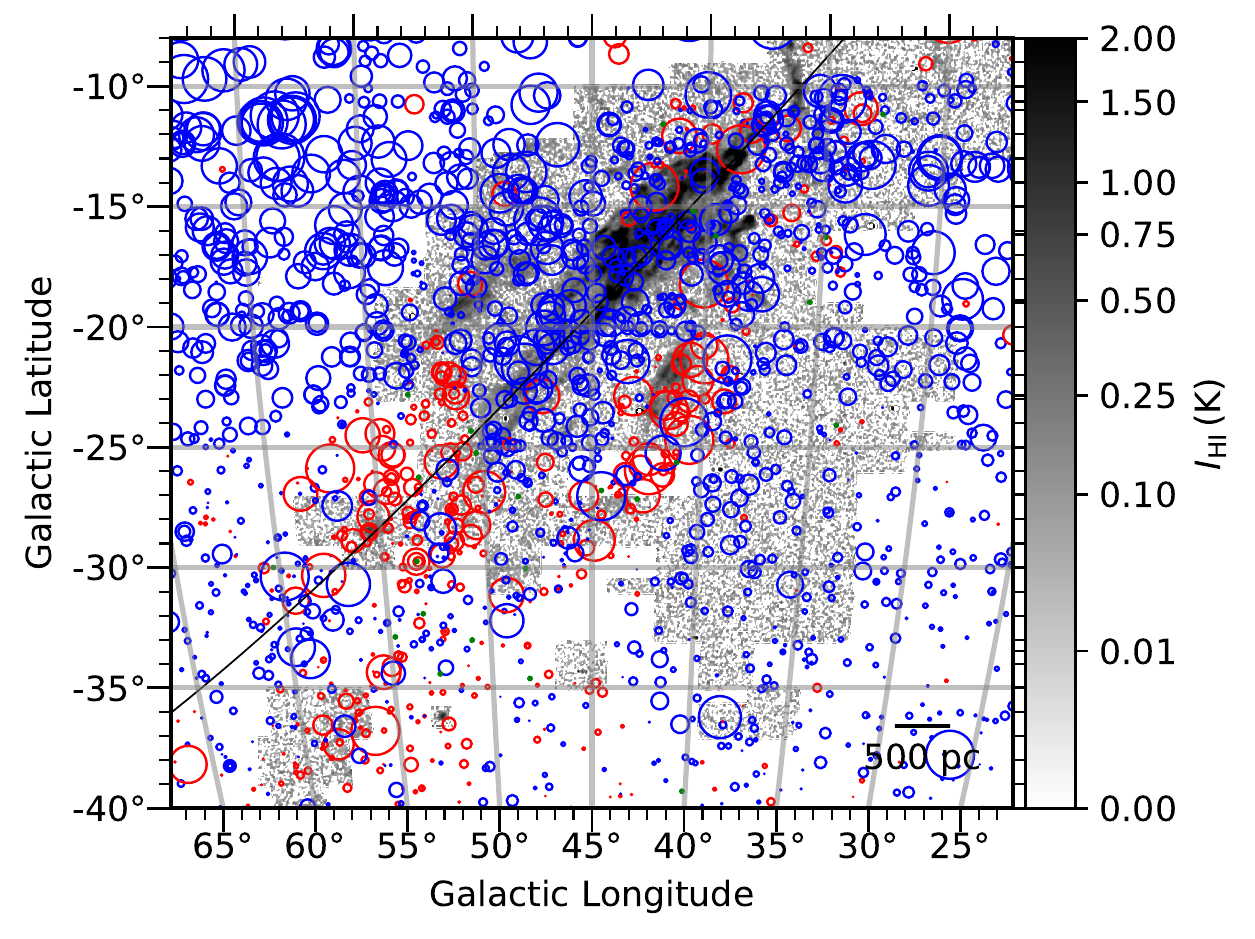}
    \end{minipage}
    \caption{Raw RMs overlaid on GBT \ion{H}{1} data (Left: JVLA RMs are bold and red/blue while \citetalias{Taylor2009} RMs are thin and magenta/cyan, Right: JVLA RMs and \citetalias{Taylor2009} RMs are the same thickness and color).  The grayscale shows \ion{H}{1} emission in the $v_{GSR} = +247$ km s$^{-1}$ channel. 
    Positive RMs (RMs $>$ 0, pointing towards the observer) are in red/magenta, negative RMs (RM $<$ 0, pointing away) are in blue/cyan, and RMs consistent with zero are in green.  The magnitude of the RM corresponds linearly to the diameter of the circle, with the largest circles representing $|\text{RM}|$ $\geq$ 200 rad m$^{-2}$.  The major axis we adopt in Figure~\ref{RMvsRADEC} below is shown with a thin black line.}
    \label{RMmaps-Small}
\end{figure*}

\begin{figure}[tb]
\centering
\begin{minipage}{.5\textwidth}
        \centering
        \includegraphics[width=\linewidth]{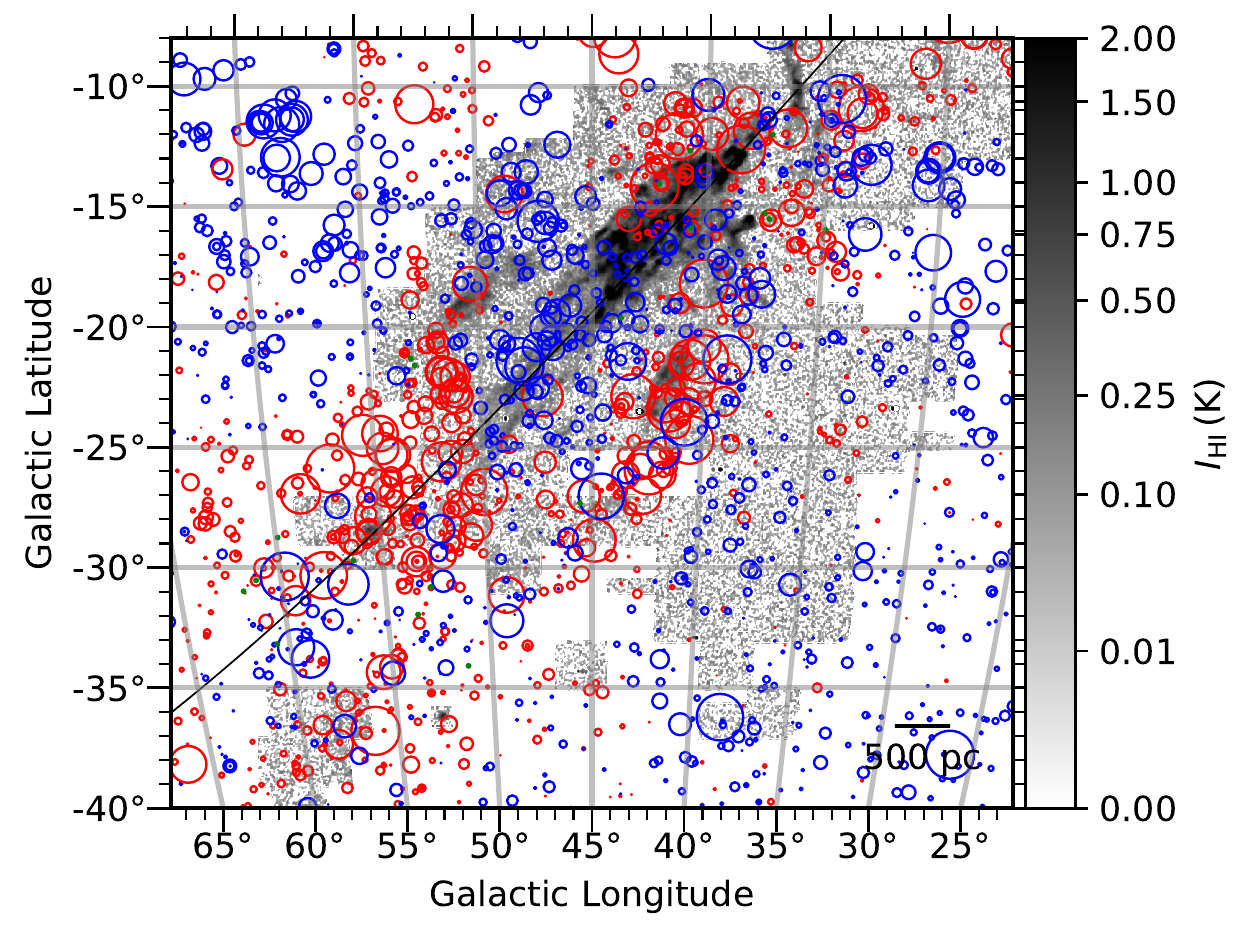}
    \end{minipage}%
\caption{Fit subtracted JVLA and \citetalias{Taylor2009} RMs (\RMhvc) overlaid on GBT \ion{H}{1} data where the grayscale shows the \ion{H}{1} emission at $v_{GSR} = +247$ km s$^{-1}$.  The foreground RM contribution from the Milky Way has been subtracted out leaving RM$_{\text{HVC}}$s. Colors and the thin black line are as in Figure \ref{RMmaps-Small}.}
\label{RMmaps-Large}
\end{figure}

\begin{figure}[tb]
 \hspace{-0.5cm}
\includegraphics[width=1.1\linewidth]{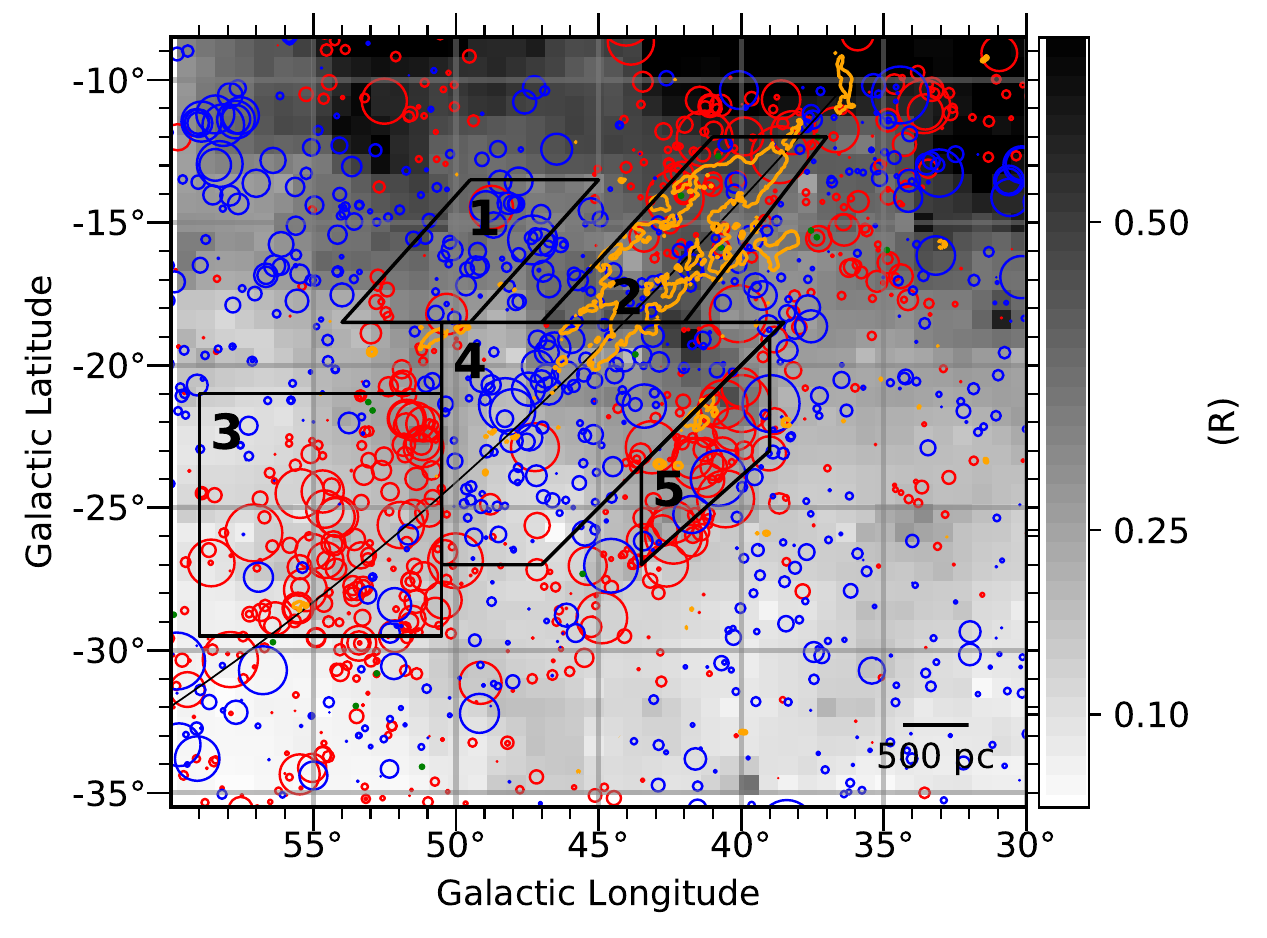}
\caption{WHAM-NSS H$\alpha$ emission measures overlaid with \RMhvc\ as in Figure~\ref{RMmaps-Large}. The 5 regions used to compute the magnetic field and an orange contour of the \ion{H}{1} emission of the Smith Cloud are shown with black polygons. The orange contour shows the GBT \ion{H}{1} emission at $v_\mathrm{GSR} = +247$ km s$^{-1}$, while the black diagonal line indicates the celestial equator. The WHAM-NSS H$\alpha$ was integrated from +25 km s$^{-1} < v_\mathrm{LSR} < + 50$ km s$^{-1}$, the Sagittarius arm contribution was subtracted out (Equation 7), and the WHAM beams smoothed.}
\label{Bfieldregions}
\end{figure}

\begin{figure}[tb]
\centering
\includegraphics[width=1.1\linewidth]{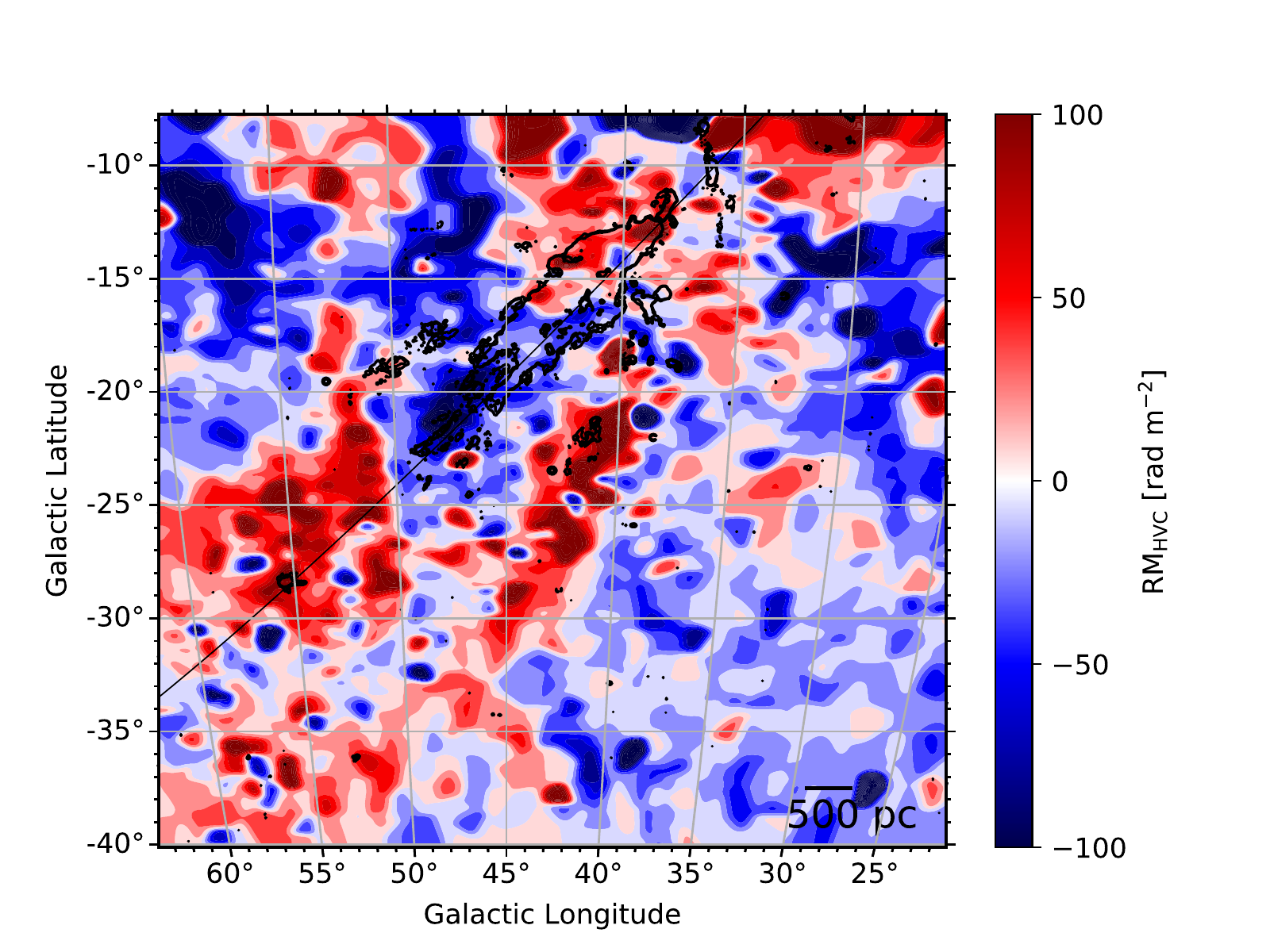}
\caption{Gaussian smoothed map of \RMhvc\ (Fig. \ref{RMmaps-Large}) overlaid with GBT \ion{H}{1} emission contour at $v_{GSR} = +247$ km s$^{-1}$.  The RM$_\mathrm{HVC}$s are smoothed with a $1.0{\arcdeg}$ kernel.  The thin black line is as in Figure \ref{RMmaps-Small}.}
\label{RMmaps-Largecontour}
\end{figure}

Using RM synthesis, we detect 1105 RMs (39$\%$ of observed sources) with a polarized intensity S/N $\geq$ 4. The resulting RMs are listed in Table \ref{first10RM}.  Of the 1105 detections, $\sim60\%$ were found with the FIND algorithm while $\sim40\%$ were found by eye.  We did not achieve the expected signal to noise with our $10-30$ second exposures, so $\gtrsim 50\%$ of our sources were not detected in polarization and total intensity.

The average uncertainty in individual RM measurements was approximately 4 rad m$^{-2}$ and the average FWHM of the RMSF was 51 rad m$^{-2}$, as we were able to recover 70$\%$ of the $\lambda^2$ space available.  Of the observed RMs, 282 had a corresponding RM from \citetalias{Taylor2009}. We compare these RMs in Figure~\ref{figure:TvsVLA}. The average difference between matches was 16.1 rad m$^{-2}$ and the median difference between matches was 9.5 rad m$^{-2}$.

We performed a Spearman rank correlation test on the matching RMs (coefficient value of 0.89) and 2D Kolmogorov-Smirnov (K-S) tests between the data sets and five regions (average coefficient value of 0.15) to test the correlation between the two data sets.  With the Spearman rank correlation test, we can reject the null hypothesis of no correlation between the data sets. The K-S test indicates that the two data sets are consistent with being drawn from the same distribution.  \citet{MaMao:2018} further explore the correlation between JVLA and \citetalias{Taylor2009} RMs.  The datasets were combined to calculate the magnetic field in order to include far more NVSS off cloud sources.  In total, we combined 2352 NVSS derived RMs with the 1105 new JVLA RMs.  In order to avoid double counting, we removed the 282 matches from the \citetalias{Taylor2009} catalog yielding 987 sources on cloud and 2188 sources off cloud (see Table \ref{RMresults}). Though the source density through the cloud is much higher in the JVLA data set, there are more total \citetalias{Taylor2009} sources because the survey covers a larger area.

In Figure \ref{RMmaps-Small}, RMs of background sources are overlaid on an \ion{H}{1} map of the Smith Cloud taken from a survey of the Smith Cloud and its environment using the 100~m Robert C. Byrd Green Bank Telescope (GBT) of the Green Bank Observatory.  The left panel differentiates between the JVLA and \citetalias{Taylor2009} RMs.  In the upper left of both panels, most RMs are large and negative (blue) while the lower right has smaller though still negative RMs.   There is a large group of positive (red) RMs in the lower left below the \ion{H}{1} emission of the Smith Cloud with a sharp break between the positive and negative RMs at $l = 50{\arcdeg}$.  

Following \citetalias{Hill2013} and \citet{McClureGriffiths2010}, we assume that the RM contribution from the foreground Milky Way can be modeled with a 2D fit.  We perform a 2D fit of the RMs as a function of $l$ and $b$ given as
\begin{equation}
\text{RM}_\text{fit} = c_0 + c_1 l + c_2 b,
\end{equation}
in order to subtract out this contribution.  We find  $c_0=-55.17 \textrm{ rad m}^{-2}$, $c_1 = -0.78 \textrm{ rad m}^{-2} \deg^{-1}$, and $c_2 = -2.86 \textrm{ rad m}^{-2} \deg^{-1}$.  We assume that the subtracted RMs, $\text{RM}_{\text{HVC}} = \text{RM} - \text{RM}_{\text{fit}}$, are RMs associated with the magnetized ionized gas of the Smith Cloud.  They are displayed in Figure \ref{RMmaps-Large} on a map of \ion{H}{1} and in Figure~\ref{Bfieldregions} on a map of H$\alpha$.  With the subtraction of the fitted foreground, the geometry associated with the cloud is highlighted and the magnetic field can be measured.  In Figure~\ref{RMmaps-Largecontour}, we show a map in which we have regridded the RM$_\mathrm{HVC}$s onto a regular grid with the RMs Gaussian-smoothed with a FWHM $1.0{\arcdeg}$ smoothing kernel.

\begin{figure}[tb]    
\centering
        \includegraphics[width=\linewidth]{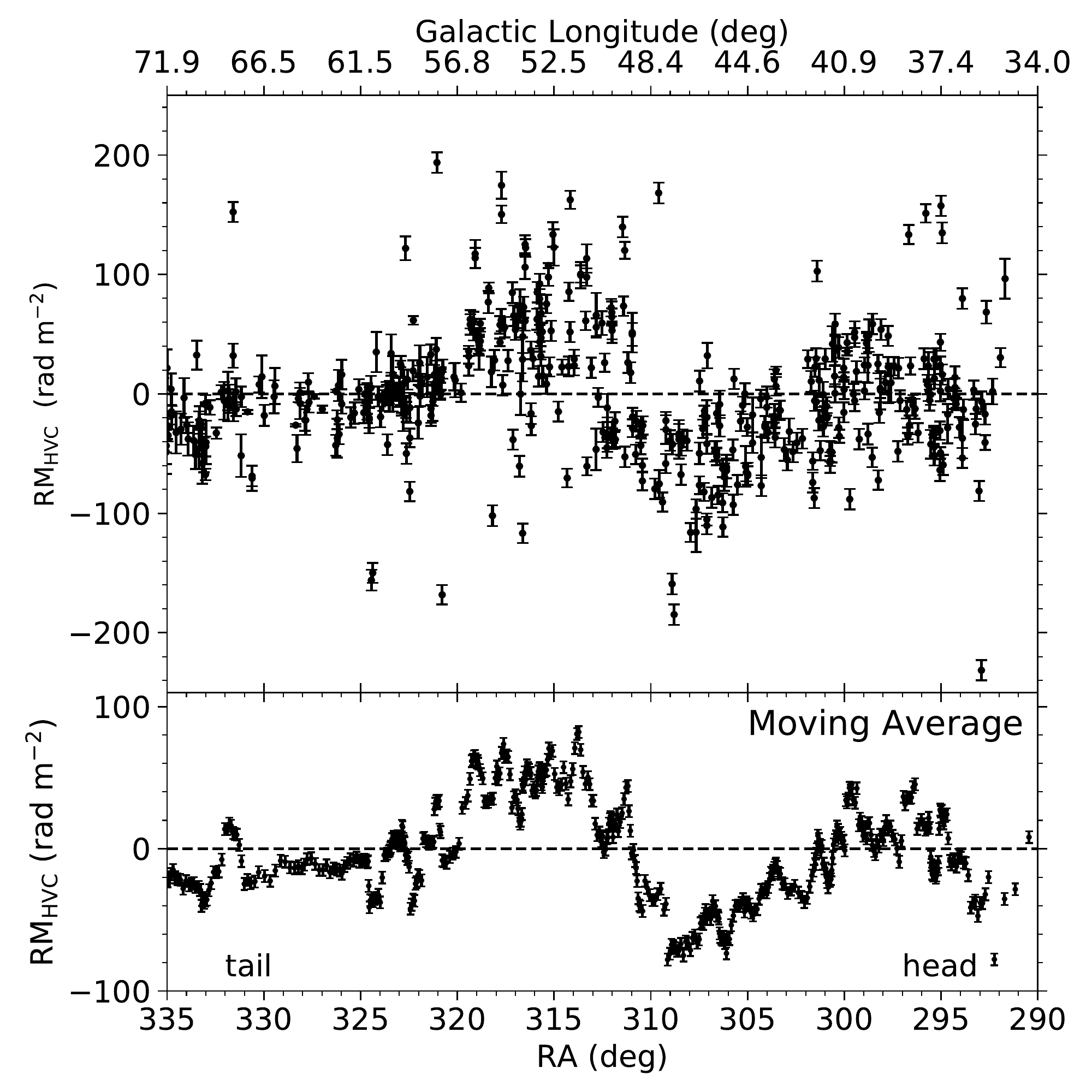}       
    \caption{RM$_{\text{HVC}}$s within $2{\arcdeg}$ of the major axis of the Smith Cloud as a function of RA.  (The major axis of the Smith Cloud lies approximately along the celestial equator.) The sign and magnitude of the RMs along the major axis show how the magnetic field changes from the head to the tail of the cloud.  The bottom panel shows a moving average with a window size of 9 RMs highlighting the sharp break at RA $= 313{\arcdeg}$ ($l = 51{\arcdeg}$).  The top axis shows the Galactic longitude corresponding to the RA along the major axis; points not exactly on the major axis have somewhat different longitudes.}
    \label{RMvsRADEC}
\end{figure}

In Figure~\ref{RMvsRADEC}, we show \RMhvc\ for points near the major axis (the black lines in Figs.~\ref{RMmaps-Small}--\ref{RMmaps-Largecontour}) of the cloud as a function of RA.
We can see how the RMs change with increasing RA.  The body of the cloud ($44\arcdeg \le l \le 50.5{\arcdeg}$) is dominated by negative RMs while the tail and head are primarily composed of positive RMs.  From Figure \ref{RMvsRADEC}, the sharp break from negative to positive RMs around $l$ = 50${\arcdeg}$ (RA $=312{\arcdeg}$) is evident. 
   
\begin{deluxetable}{ccccc}[b]
\tablewidth{0pt}  
\tablecaption{Number of Detected RMs \label{RMresults}
}
\tablehead{\colhead{} & \colhead{JVLA } & \colhead{\citetalias{Taylor2009}}  & \colhead{Matches} & \colhead{Total}}
\startdata
On Smith Cloud & 854  & 354 & 221& 987 \\
Off Smith Cloud & 251 & 1998  & 61  & 2188\\
\hline
Total & 1105 & 2352 & 282 & 3175 \\
\enddata
\tablecomments{On Smith Cloud refers to sources within $\pm10{\arcdeg}$ of the major axis while off Smith Cloud refers to all other sources between $\pm10{\arcdeg}$ and $\pm20{\arcdeg}$ of the major axis.  The total amount of sources on and off cloud is found by removing the matching sources from the \citetalias{Taylor2009} catalog.}
\end{deluxetable}

%------------------------------------------------------------------------------------------------------------------------- 
 
\section{Magnetic Field Strength Estimation} \label{Sec4}
We calculate the magnetic field strength in five regions of the Smith Cloud following the method used by \citetalias{Hill2013}.  We chose regions $1,3,4$ and $5$ defined in Figure \ref{Bfieldregions} such that the RMs enclosed are predominately positive or predominately negative.  We selected region $2$ to encompass the head and main body of the cloud. 
To measure the magnetic field, we estimate the electron density associated with the Fardaday rotating gas. \citetalias{Hill2013} (see their Figure 5) argued that the Faraday rotation occurs in both decelerated ($v_\mathrm{LSR} \approx +40 \textrm{ km s}^{-1}$) gas and gas at the Smith Cloud velocity ($v_\mathrm{LSR} \approx +100 \textrm{ km s}^{-1}$ near the tip). We derive H$\alpha$ emission measures
\begin{equation} \label{eqn7}
\begin{aligned}
\text{EM} {} & \equiv \int n_{e}^{2} (s)ds \\
	{} & = 2.75\left(\frac{T}{10^{4}\text{ K}}\right)^{0.9} \frac{I_\mathrm{SC}}{\text{R}} \text{ pc cm}^{-6},
\end{aligned}
\end{equation}
where $I_\mathrm{SC}$ is the H$\alpha$ intensity due to the Smith Cloud, Rayleigh (R) is the unit of flux, and we assume $T = 8000$~K in H$\alpha$-emitting gas (see \citetalias{Hill2013} for a full discussion of this derivation).

We find the average EM for five regions of the Smith Cloud from the average H$\alpha$ intensity from WHAM observations. Regions~1 and 5 are morphologically associated with the decelerated H$\alpha$ ridge, whereas Smith Cloud velocity gas is most likely responsible for the Faraday rotation along the main axis of the cloud (regions 2 and 4), while the situation in region 3 is less clear \citepalias{Hill2013}. We therefore use $+25 \textrm{ km s}^{-1} < v_\mathrm{LSR} < +50 \textrm{ km s}^{-1}$ H$\alpha$ emission from the WHAM Northern Sky Survey \citep[WHAM-NSS,][]{Haffner2003} to estimate the EM in regions 1 and 5, and the $v_\mathrm{LSR} \approx +100 \textrm{ km s}^{-1}$ H$\alpha$ from the \citet{Hill2009} WHAM observations in regions 2 and 4. We do not have Smith Cloud-velocity H$\alpha$ observations which cover region 3; we only have the WHAM-NSS data. In region 3, the \ion{H}{1} emission of the cloud is mostly found at $v_\mathrm{LSR} \approx +70 \textrm{ km s}^{-1}$, at the edge of the WHAM-NSS velocity range; there is no clear detection of H$\alpha$ emission at these velocities. However, the $l=51{\arcdeg}$ ridge (Figure 5 of \citetalias{Hill2013}) is detected in both \ion{H}{1} and H$\alpha$ at $v_\mathrm{LSR} \approx +40 \textrm{ km s}^{-1}$. Therefore, we use the $+25 \textrm{ km s}^{-1} < v_\mathrm{LSR} < +50 \textrm{ km s}^{-1}$ EM in region 3. We list these EMs in Table~\ref{Bfieldtable}.

Combining equations \ref{eqn2} and \ref{eqn7} and with the assumption that the magnetic field does not vary with RM or path length, the average line-of-sight magnetic field is  
 \begin{equation}
\langle B_{||} \rangle =\frac{\langle \text{RM}_{\text{HVC}} \rangle}{0.81 \times (L_{H^{+}} \langle \text{EM} \rangle)^{1/2}}.
\end{equation}
$\langle\text{RM}_{\mathrm{HVC}} \rangle$ is the weighted mean of all RM$_{\mathrm{HVC}}$s in each region in Figure \ref{Bfieldregions}.  
The weights are $w_i=\sigma_{i}^{-2}/(\sum_{i}\sigma_{i}^{-2})$ where $\sigma_{i}^2 = (\text{uncertainty in RM}_{\text{HVC},i})^{2} + (7 \text{ rad m}^{-2})^{2}$, assuming a 7 rad m$^{-2}$ standard deviation of the intrinsic RMs of the sources \citep{Schnitzeler2010, Stil2011}. 
  
Estimating the density from the EM requires assuming a path length $L_{H^{+}}$. We assume that the gas has $n_e(s) = n_e$ in a fraction $f = L_{H^{+}}/L$ of the path and 0 elsewhere.  Following \citetalias{Hill2013}, we pick $L$ for each region to be the largest value reasonable for the morphology and assume $f=1$ so that the magnetic field estimates are lower limits.  Note that the magnetic field estimates will increase as $f^{-0.5}$.  Along the H$\alpha$ ridge, the brightest H$\alpha$ emission is one WHAM beam wide; we assume that $L_{H^{+}}<$ 220 pc, the size of one beam.  For regions above and below the ridge, we assume the H$\alpha$ emission is larger and estimate a higher path length.  The resulting magnetic fields in each region are listed in Table \ref{Bfieldtable}.  

\begin{deluxetable*}{cccr@{ $\pm$ }lccr@{ $\pm$ }l}[tb]
\tablecolumns{9} 
\tablewidth{0pt}  
\tablecaption{Magnetic Field Estimates\label{Bfieldtable}
}
\tablehead{\colhead{Region} & \colhead{$l$} & \colhead{$b$} &  \multicolumn2c{$\langle\text{RM}_{\text{HVC}} \rangle$}  &  \colhead{$\langle\text{EM}\rangle$} & \colhead{$L_{H^{+}}$} & \multicolumn2c{$\langle B_{||} \rangle$}  \\ \colhead{} & \colhead{(deg)} & \colhead{(deg)} & \multicolumn2c{(rad m$^{-2}$)}  & \colhead{(pc cm$^{-6}$)} & \colhead{(pc)} & \multicolumn2c{($\mu$G)}}
\decimals
\startdata
1 & 49 & $-$16    & $-$30			& 2	 	& 1.71 & 220  	& $-$1.91 		& 0.1 \\
2 & 42 & $-$15    &  $+$6		  	&  1	  	& 0.88 & 220 	& 0.55 		& 0.2\\ 
3 & 55 & $-$25    &  $+$47			&  1	 	& 0.48 & 880  	& 2.86 		& 0.3\\
4 & 47 & $-$22    &  $-$20			&  1	  	& 0.31 & 1100   &  $-$1.35	& 0.2\\
5 & 41 & $-$22    &  	$+$72 			&  2		& 1.26 & 220   	&  5.33		& 0.3 \\
\enddata
\hspace{4cm}
\tablecomments{Uncertainties in $\langle \mathrm{RM}_\mathrm{HVC} \rangle$ are the standard deviation of $\langle \mathrm{RM}_\mathrm{HVC} \rangle$ in each region.  $L_{H^{+}}$ is the largest reasonable value for the morphology, so $\langle B_{||} \rangle$ is a lower limit.}
\end{deluxetable*}   

%-----------------------------------------------------------------------------------------------------------------------------
\section{Discussion} \label{Sec5}

With the combined data, the RMs around the Smith Cloud are sampled with 3 point sources deg$^{-2}$ on cloud and 1 point source deg$^{-2}$ off cloud. Overall, the morphology of the RMs is similar to that identified by \citetalias{Hill2013}. With the higher density of RMs of extragalactic sources through the Smith Cloud, we are able to better isolate morphological features.

\citetalias{Hill2013} identified an \ion{H}{1} emission ridge at $v_\mathrm{LSR} \approx +40$ km s$^{-1}$ at $l = 51{\arcdeg}$.  This ridge starts at $(l,b) =  (51{\arcdeg}, -21.5{\arcdeg})$, extending to $(l,b) =  (51{\arcdeg}, -29{\arcdeg})$ with a width of approximately one degree.  While this ridge was seen with RMs from just \citetalias{Taylor2009} data, the ridge is prominently defined with the combined datasets.  Along the right and top of the ridge, the RMs are negative, as seen in blue circles, while the RMs along and to the left of this ridge are positive.  The ridge can also be seen in Figure \ref{RMvsRADEC} at $\mathrm{RA} = 313{\arcdeg}$ by looking at the location of the RMs along the major axis of the Smith Cloud.  As the major axis of the cloud lies along the celestial equator, by looking at RM as a function of RA, we see how the RMs change from the head to the tail of the cloud.  From $\mathrm{RA} \approx 309{\arcdeg}$, the RMs decrease in magnitude with increasing RA, indicating a field pointing away from the sun until $\mathrm{RA} \approx 312{\arcdeg}$ (corresponding to $l = 50{\arcdeg}$ at the equator), after which the RMs are predominately positive. This sign change in the RM at $l = 51{\arcdeg}$ is also evident in the maps of both raw RMs (Figure \ref{RMmaps-Small}) and foreground-subtracted RMs (Figures \ref{RMmaps-Large} and \ref{Bfieldregions}). This indicates that the magnetic field changes sharply at the \ion{H}{1} ridge and then remains positive to the left of the body of the Smith Cloud.  As we see no corresponding change in either H$\alpha$ at $v_\mathrm{LSR} = +40 \textrm{ km s}^{-1}$ or in the \ion{H}{1} emission, which in Figure \ref{RMmaps-Large} is shown at Galactic standard of rest (GSR), $v_\mathrm{GSR} = +247 \textrm{ km s}^{-1}$, this sign change in the magnetic field is most likely due to a foreground feature or a feature only seen at a different \ion{H}{1} GSR velocity.

In the models presented by \citet{Gronnow2017}, there is an enhancement in the field in the tail relative to the head and especially the body. In the Smith Cloud tail (region~3), we have $|B_{||}|\gtrsim+2.9 \, \mu \mathrm{G}$, compared to $|B_{||}\lesssim-1.9 \, \mu \mathrm{G}$ in the head (region~1, Table~\ref{Bfieldtable}). Therefore, if the field in region~3 is truly associated with the Smith Cloud and not a foreground feature, this is consistent with the models of \citet{Gronnow2017}. Although it is not obvious how such an enhancement could lead to a sign change such as we see, this morphology is a potential avenue for direct comparison of the observations to the predictions of MHD simulations.

The magnetic field lower limits found in each region are consistent with the values found by \citetalias{Hill2013}.  This peak value is consistent with the  \citetalias{Hill2013} value of $+8 \, \mu$G. Our data strengthen their finding that the RMs along the narrow, decelerated H$\alpha$ ridge in region 5 are much stronger and opposite in sign compared to the surrounding RMs, suggesting a compressed ambient field which has a preferred direction.  

Due to the high density of RMs on and next to the Smith Cloud, we can model the morphology of the magnetic field.  As seen in Figures \ref{RMmaps-Large} and \ref{Bfieldregions}, an arch of positive RMs extends above and around the head of the Smith Cloud.  This cap of RMs suggests a magnetic field that has been compressed by the ambient field as the Smith Cloud travels towards the Galactic plane assuming that the cloud is falling into the disk (see Section \ref{Sec1}).  

Under the assumption that $|\vec{B}|$ is roughly constant, since RM is proportional to $B_{||}$, the small negative RMs over the body suggest that the perpendicular component of the magnetic field is stronger than the parallel component, while the large positive RMs surrounding the cloud indicate that the parallel component dominates.  To the right of the major axis of the cloud, the RMs are predominately positive, indicating a weak magnetic field perpendicular vector component pointing towards the observer.  To the left of the Smith Cloud, strong negative RMs dominate.  These strongly negative RMs indicate weak magnetic field perpendicular vector components pointing away from us.  This picture is qualitatively consistent with strong parallel magnetic field vector lines along the sides of the Smith Cloud.  This suggests a magnetic field that is draped, or laid over the ionized gas of the Smith Cloud, where the field has been compressed near the tip of the cloud \citep{Dursi2008}.    

This configuration of a draped magnetic field surrounding the Smith Cloud is consistent with the idea that the neutral gas in the Smith Cloud is shielded from the hot ISM by a magnetic barrier between the surrounding medium and the gas; this has been seen in 2D simulations by \citet{Konz2002} and in 3D simulations by \citet{Gronnow2017}.  \citet{Gronnow2017} found that a magnetic field of a few $\mu$G associated with an HVC within 10 kpc of the disk corresponds to an ambient field that has been ``swept up" by the cloud, which is consistent with \citetalias{Hill2013} and our observations.  The \citet{Gronnow2017,Gronnow2018} models also support the idea that magnetic field could keep the cloud intact as it travels through the Galactic halo by delaying hydrodynamic instabilities and stripping of the gas.    
With this draped magnetic field barrier, the Smith Cloud could travel through the ISM keeping the gas near the head and body intact.
To test if this geometry could produce the observed RMs, we create a toy model of a possible magnetic field configuration surrounding the Smith Cloud. 

%-----------------------------------------------------------------------------------------------------------------------------
\subsection{Magnetic Field Model of the Smith Cloud}

In order to visualize a possible configuration of the magnetic field inferred from the foreground subtracted RMs ($\langle$RM$_{\text{HVC}}\rangle$), we create a 3D magnetic field model to determine how the electron density and magnetic field could produce the RMs seen in observations.  We surround an ellipsoidal cloud (axes: $3\times1\times1$ kpc$^{3}$) with a thin ionized shell of constant thickness.  The cloud is set at a $45{\arcdeg}$ angle such that the disk of the galaxy would be overhead and the cloud would be falling into a transverse magnetic field plane (The cloud has projected axes ($i$, $j$, $k$), where $i$ is along the minor axis, $j$ is into the page, and $k$ is along the major axis).  The cloud, ionized shell, and surrounding medium each have a uniform electron density within each component.  

\begin{figure}[tb]
  \includegraphics[width=\linewidth]{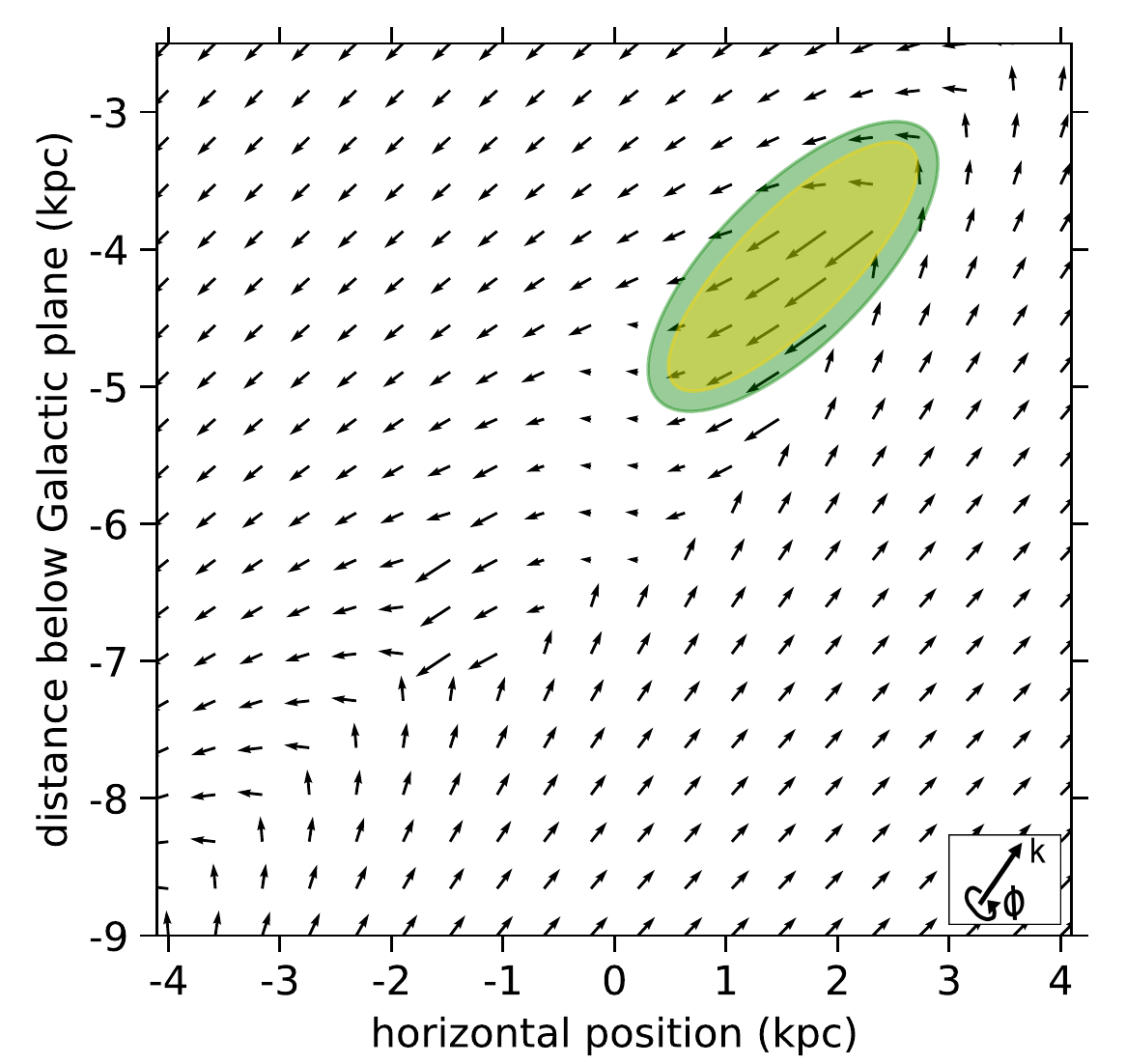}
  \caption{Model set up for the Smith Cloud at the center slice.  The gold ellipsoid represents the \ion{H}{1} emission at the head of the Smith Cloud and the green represents the ionized shell.  The arrows represent the magnetic field lines.}
  \label{RMmodel}
\end{figure}

We create a $8\times 8\times 8$ kpc$^{3}$ cell grid, where the neutral and ionized gas density and the magnetic field strength could be changed to match observations.  We use an electron density of 0.01 cm$^{-3}$ for the surrounding medium, the ionized skin has an electron density of 0.09 cm$^{-3}$, and the cloud is 0.04 cm$^{-3}$ \citep{Hill2009}.  Along the edges of the grid, the electron density drops to 0 cm$^{-3}$. The magnetic field is modeled to produce field lines that drape over the cloud as suggested in MHD simulations of a cloud falling onto planar field lines \citep{Konz2002,Romanelli2014,GalyardtShelton:2016}. We set
\begin{align} 
    B_{i} &= B_{0} e^{-r/h_r}  \cos(\phi) \label{Bx} \\ 
    B_{j} &= 0 \\
    B_{k} &= \pm B_{0} (1-e^{-2r/h_r})^{1/2}. \label{Bz}
\end{align}

Here $r$ and $\phi$ are the cylindrical coordinates measured with respect to the major axis of the cloud.  As this is a toy model, we do not construct a vector potential field and as such, $\nabla \cdot \vec{B}$ is not always zero.  Inside the cloud and ionized shell, we follow \citet{Gourgouliatos2010} and assume toroidal and poloidal magnetic fields.  Figure \ref{RMmodel} shows the model setup with a slice through the center of the cloud.

\begin{figure}[tb]
  \includegraphics[width=\linewidth]{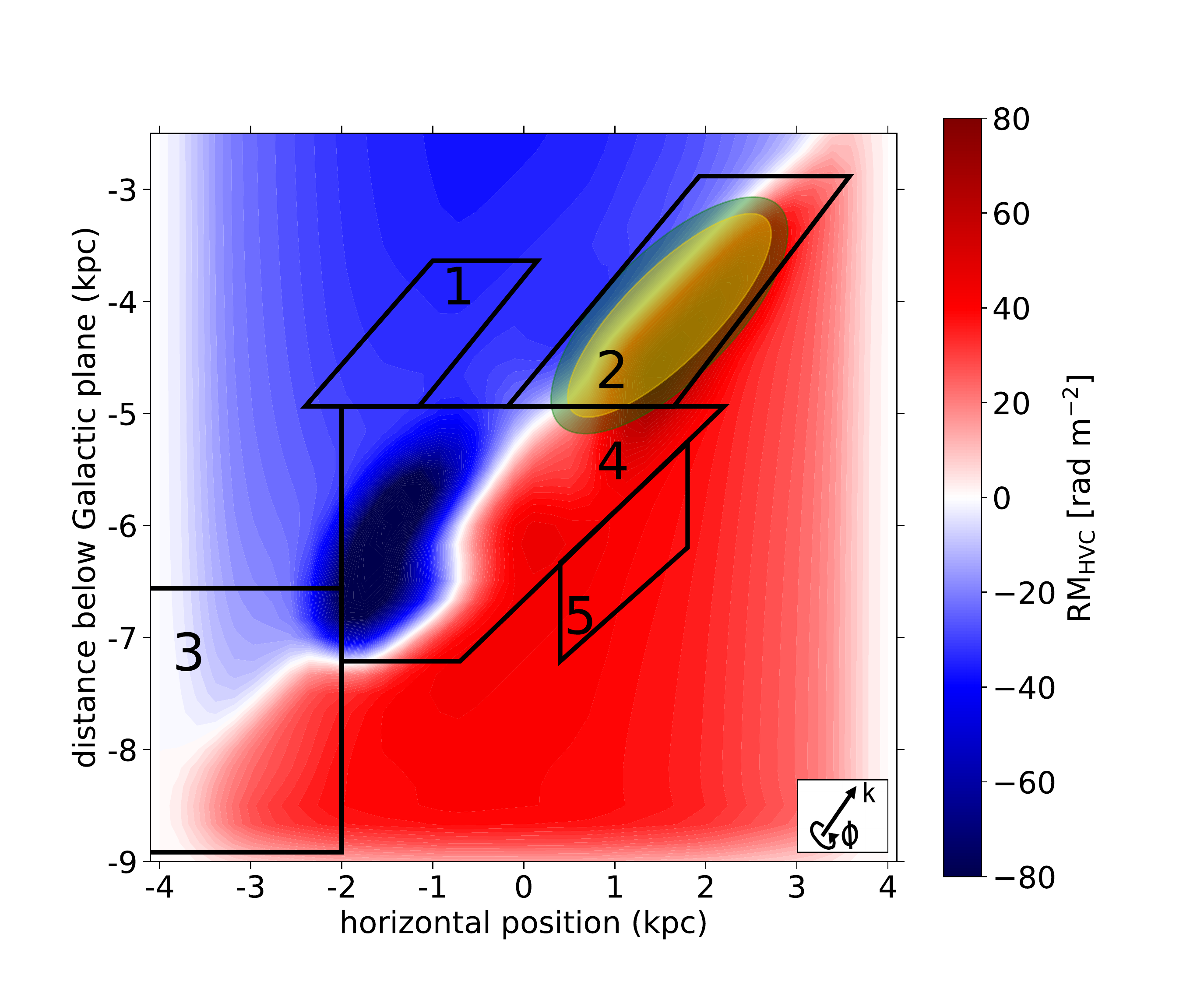}
  \caption{RMs produced by the 3D model for an initial $B_{0} = +5$ $\mu$G and $\theta = -4{\arcdeg}$ (run $\#$7).  The gold ellipse represents the \ion{H}{1} emission at the head of the Smith Cloud and the green represents the ionized shell.}
  \label{RMregions}
\end{figure}

As the Smith Cloud lies below the Galactic plane  
at $b = -13{\arcdeg}$, observed RM measurements are determined by assuming we are viewing the magnetic field at this latitude.  The magnetic field is then tilted by some angle ($\theta$) with respect to the Galactic plane in order to vary the distance to the head or tail.  (The head is taken to be at a horizontal position of 3 kpc with the tail streaming behind until $-1$ kpc.) We ran models with a range of values of $B_0$ and $\theta$, listed in Table~\ref{Valuesmodel}. Figure~\ref{RMregions} shows the resulting synthetic RMs derived from our preferred model.

For each initial magnetic field strength and $\theta$, we calculate the line -of-sight (parallel) component of the magnetic field and the electron density at every point along the line of sight from equations \ref{Bx}--\ref{Bz}. The rotation of the plane in which the model field lies introduces the line-of-sight component of the field.  This component of the magnetic field is taken to be the projection of each component along the sightline.  RMs are then calculated following equation \ref{eqn2}.  

To test the model, we vary the strength of the magnetic field and $\theta$ through the plane.  As we did not vary the electron density, the results from the model are electron density dependent and affect RMs found for each run.  The strength of the magnetic field varies from $-10$ to $+10 \, \mu$G and $\theta$ varies from $-7$ to $7$ degrees.  Changing the strength of magnetic field and $\theta$ affects the magnitude and geometry of the RMs (Table \ref{Valuesmodel}). 

As this model only accounts for magnetic field strength and electron density, we focus on the broad pattern of RMs on and directly next to the Smith Cloud. 
The enhancement of positive RMs at $l=51{\arcdeg}$ seen in observations is neglected as we are only modeling the draped magnetic field.  (We could of course make the model fit by adding a positive-field component. One way to do that would be a twist in the cloud.)
Therefore, model RMs from region 3 will not match the observations in the same region.

From Table \ref{Valuesmodel}, runs 5--9 (including run 7, shown in Figure~\ref{RMregions}) produce RM patterns in Region 1 and 5 with the best qualitative match to the observations.  The draped magnetic field produces negative RMs to the left of the cloud and positive RMs to the right of the cloud.  These runs keep the RMs within the range of observed RMs and around the mean RM value observed in each region.  As the model assumes constant densities and $B_y = 0$ along with a toriodal field within the cloud, by tilting the field, the mean RMs in each region increase in magnitude in increasing tilt.

Although the shell surrounding the body of the modeled cloud had a different density from the body, this did not significantly affect the RM strength or direction.  In region 5, the observed $\langle$RM$_{\text{HVC}}\rangle = +72 \pm 2$ rad m$^{-3}$ and the
modeled average RM of the region for runs with $B_0 = 5 \ \mu\rm{G}$ and $-2\arcdeg < \theta < -7$ is RM $= +48$ rad m$^{-2}$.  While in region 1, the observed $\langle$RM$_{\text{HVC}}\rangle = -30 \pm 2$ rad m$^{-2}$ and the modeled average RM of the region for runs with $B_0 = 5 \ \mu\rm{G}$ and $\theta$ between $-3{\arcdeg}$ to $-7{\arcdeg}$ is RM $= -39$ rad m$^{-2}$. 
   
\begin{figure}[tb]
\centering
\includegraphics[width=1.1\linewidth]{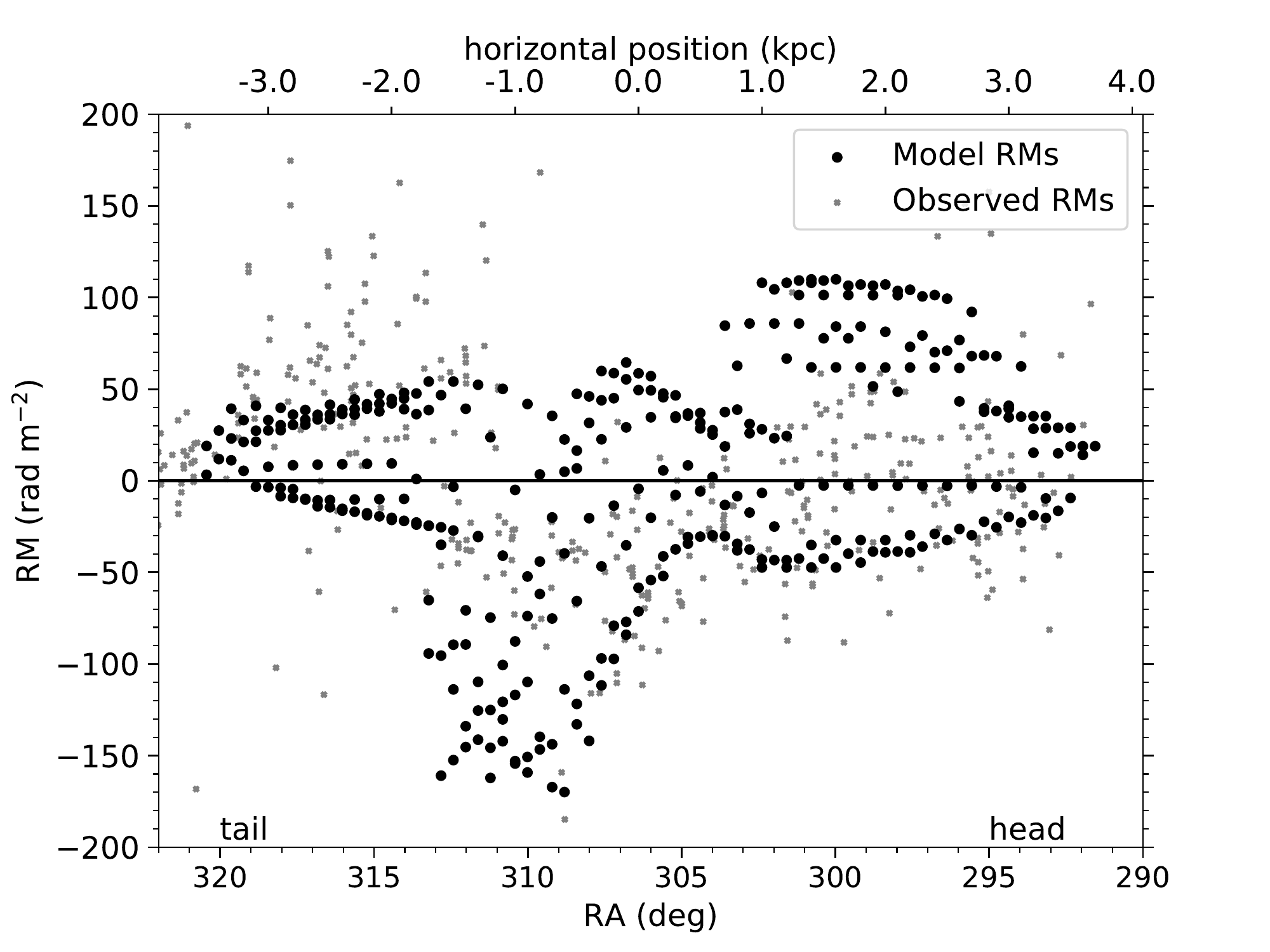}
\caption{Comparison of RMs from the model (black dots) to \RMhvc\ (grey dots) near the major axis of the Smith Cloud (Fig.~\ref{RMvsRADEC}).}
\label{modelRMvsRM}
\end{figure}

Following along the major axis of the modeled cloud in a range of $\pm 1$ kpc, the RM profile for run $\#$7 aligns with the RMs found in observations (Figure \ref{modelRMvsRM}; excluding the tail where we did not include the $l=51{\arcdeg}$ feature).  The grey points are the observed RMs shown in the top panel of Figure \ref{RMvsRADEC}. 

From this toy model, we are able to produce RM maps similar to the Milky Way foreground subtracted RMs seen in region 1 and 5 of Figure \ref{RMmaps-Large}, indicating that this is a plausible though non-unique picture of the geometry of the magnetic field of the Smith Cloud. 

\begin{deluxetable*}{ccc|cccccccc}
\tablewidth{0pt}  
\tablecaption{Model Ranges in RM \label{Valuesmodel} 
} 
\tablehead{\multicolumn{3}{c}{Inputs}  &  \multicolumn{8}{c}{Outputs} }
%\centering
\startdata	
Run & $B_{0}$ & $\theta$ & Min RM & Max RM &  Mean RM & Region 1 & Region 2 & Region 3 & Region 4 & Region 5  \\ 
 & ($\mu$G) & (deg)  & (rad m$^{-2}$)  & (rad m$^{-2}$) & (rad m$^{-2}$) & (rad m$^{-2}$) & (rad m$^{-2}$) & (rad m$^{-2}$) & (rad m$^{-2}$) & (rad m$^{-2}$) \\ %[.4ex]
 \hline
 1	&	$-$10	&	$-$4	&	$-$186.7	&	$+$74.8	&	$-$18.3	&	$+$65.8	&	$-$48.8	&	$-$29.5	&	$-$61.5	&	$-$78.4	\\
2	&	$-$1	&	$-$4	&	$-$133.8	&	$+$10.5	&	$-$4.7	&	$+$6.6	&	$-$5.7	&	$-$7.9	&	$-$24.5	&	$-$7.8	\\
3	&	1	&	$-$4	&	$-$123.3	&	$+$18.9	&	$-$1.7	&	$-$6.6	&	$+$3.9	&	$-$3.1	&	$-$16.2	&	$+$7.8	\\
4	&	5	&	$-$1	&	$-$26.8	&	$+$34.0	&	$+$1.6	&	$-$8.2	&	$+$3.0	&	$+$1.2	&	$+$3.6	&	$+$9.9	\\
5	&	5	&	$-$2	&	$-$64.5	&	$+$43.3	&	$+$2.3	&	$-$16.7	&	$+$8.7	&	$+$2.9	&	$-$2.0	&	$+$20.3	\\
6	&	5	&	$-$3	&	$-$105.2	&	$+$66.0	&	$+$3.3	&	$-$25.7	&	$+$15.3	&	$+$4.7	&	$-$6.3	&	$+$31.3	\\
7	&	5	&	$-$4	&	$-$139.7	&	$+$86.6	&	$+$4.4	&	$-$33.0	&	$+$23.1	&	$+$6.8	&	$-$10.5	&	$+$40.4	\\
8	&	5	&	$-$5	&	$-$169.9	&	$+$109.9	&	$+$5.7	&	$-$42.3	&	$+$33.3	&	$+$9.5	&	$-$13.6	&	$+$51.8	\\
9	&	5	&	$-$7	&	$-$257.5	&	$+$151.6	&	$+$8.9	&	$-$56.9	&	$+$56.1	&	$+$15.5	&	$-$18.2	&	$+$70.3	\\
10	&	5	&	1	&	$-$43.3	&	$+$158.4	&	$+$0.5	&	$+$16.9	&	$-$1.2	&	$+$4.5	&	$+$15.5	&	$-$20.0	\\
11	&	5	&	3	&	$-$66.0	&	$+$205.6	&	$+$0.3	&	$+$26.2	&	$+$0.1	&	$+$6.0	&	$+$20.5	&	$-$31.0	\\
12	&	5	&	4	&	$-$86.8	&	$+$244.5	&	$+$0.5	&	$+$33.9	&	$+$2.6	&	$+$8.7	&	$+$26.1	&	$-$40.2	\\
13	&	5	&	5	&	$-$109.7	&	$+$297.0	&	$+$0.7	&	$+$43.8	&	$+$9.9	&	$+$12.5	&	$+$29.0	&	$-$51.6	\\
14	&	5	&	7	&	$-$151.3	&	$+$374.3	&	$+$2.2	&	$+$60.1	&	$+$22.5	&	$+$20.8	&	$+$41.7	&	$-$71.9	\\
15	&	10	&	4	&	$-$163.9	&	$+$173.2	&	$+$11.9	&	$-$65.8	&	$+$47.1	&	$+$18.6	&	$+$20.8	&	$+$78.4	\\
\enddata
\end{deluxetable*}

%-----------------------------------------------------------------------------------------------------------------------------
\section{Summary} \label{Sec6}
In this paper we have presented RMs for 1105 polarized extragalactic sources behind and next to the Smith Cloud as listed in Table \ref{first10RM}.  In Section \ref{Sec3}, we have confirmed a correlation between these RMs and RMs found from \citetalias{Taylor2009} using the NVSS survey and have combined the two datasets giving 3175 RMs on and surrounding the cloud with 3 sources per square degree through the cloud.  The combined RMs allow us to robustly measure the morphology of the magnetic field of the Smith Cloud.  The detailed RM map affirms previously found aspects of the cloud from \citetalias{Hill2013}, including the sharp \ion{H}{1} ridge along $l = 51{\arcdeg}$ seen in a sharp line of positive RMs \citep{Lockman2008}.    

In Section \ref{Sec4}, after correcting for the foreground Milky Way contribution, we use the RMs, along with H$\alpha$ upper limit emission measures and upper limit path lengths to estimate a line-of-sight magnetic field strength of $+5~\mu\mathrm{G}$.  This magnetic field value is consistent with the result found by \citetalias{Hill2013}.  From the geometry of the RMs surrounding the Smith Cloud and the magnetic field values, we suggest a strong magnetic field has formed around the cloud from the compressed ambient field through which the Smith Cloud travels.
To determine if a magnetic field draped, or laid over the cloud could produce the observed RMs and the magnetic field strength, we create a non-unique model of the magnetic field of the Smith Cloud.  We show that the average RM measurements are consistent with observations, and the overarching geometry of the Smith Cloud can be produced with a magnetic field strength of +5 $\mu$G observed below the disk of the Milky Way.      

Further modeling and simulations of this geometry and field strength will confirm if the Smith Cloud can survive its full passage without being stripped of its gases.  In particular, we do not distinguish between origin scenarios for the field, which could be intrinsic to the Smith Cloud or a swept-up field which originated in the Galactic ISM.   

\acknowledgments
We thank Rainer Beck for a close reading and many thoughtful comments of this manuscript. 
S.K.B. and A.S.H. acknowledge support from NASA grant HST-AR-14297.  S.K.B acknowledges the support of Haverford College through the KINSC Summer Scholars program and the Frances Velay Research Fellowship. 
The Dunlap Institute is funded through an endowment established by the David Dunlap family and the University of Toronto. B.M.G. acknowledges the support of the Natural Sciences and Engineering Research Council of Canada (NSERC) through grant RGPIN-2015-05948, and of the Canada Research Chairs program.  N.M.M.-G. acknowledges the support of the Australian Research Council through Future Fellowship FT150100024.  The Green Bank Observatory is a facility of the National Science Foundation, operated under a cooperative agreement by Associated Universities, Inc.  The National Radio Astronomy Observatory is a facility of the National Science Foundation operated under cooperative agreement by Associated Universities, Inc.
The Wisconsin H-Alpha Mapper is supported by the National Science Foundation.    
\facilities{JVLA, GBT, WHAM}
\software{astropy \citep{Astropy-CollaborationRobitaille:2013}}, APLpy,\footnote{http://aplpy.github.io} CASA \citet{McMullin2007}, MIRIAD \citet{MIRIAD}

\singlespace

\bibliography{reference}

\end{document}